\pdfoutput=1
\documentclass[10pt,tightenlines,eqsecnum,floats,aps,amsmath,
amssymb,nofootinbib,prd,showpacs,notitlepage]{revtex4-1}

\usepackage[utf8]{inputenc}
\usepackage[OT4]{fontenc}

\usepackage{amsmath,amsthm,latexsym,amssymb,amsfonts}
\usepackage{enumerate}
\usepackage{graphicx}
\usepackage{calc}
\usepackage{colordvi}
\usepackage{color}
\usepackage{vmargin}
\usepackage{array}
\usepackage{subfig}

\newcommand{\braket}[1]{\left\langle{#1}\right\rangle}

\newcommand{\inn}{\mathrm{in}}
\newcommand{\out}{\mathrm{out}}

\newcommand{\tr}{\mathrm{Tr}}
\newcommand{\inv}{\mathrm{Inv}}

\newcommand{\pd}{\partial}

\newcommand{\K}[1]{\mathbb{#1}}	
\newcommand{\T}[1]{\mathcal{#1}}	

\newcommand{\Hil}{\mathcal{H}}

\newcommand{\tab}{\qquad\qquad}
\newcommand{\taab}{\tab\tab}

\newtheorem{lm}{Lemma}
\newtheorem{thm}[lm]{Theorem}
\newtheorem{df}{Definition}

\newcommand{\reef}[1]{(\ref{eq:#1})} 
\newcommand{\fref}[1]{Fig.~\ref{fig:#1}} 
\newcommand{\sref}[1]{Sec.~\ref{sc:#1}} 

\def\be{\begin{equation}}
\def\ee{\end{equation}}
\def\ba{\begin{eqnarray}}
\def\ea{\end{eqnarray}}

\newcommand{\s}[1]{{#1}^{(s)}}

\begin{document}


\title{Feynman diagrammatic approach to spin foams}
\author{Marcin Kisielowski${}^{1,2}$, Jerzy Lewandowski${}^{1}$, Jacek Puchta${}^{1,3}$}

\affiliation{${}^1$ Instytut Fizyki Teoretycznej, Uniwersytet Warszawski,
ul. Ho{\.z}a 69, 00-681 Warszawa (Warsaw), Polska (Poland)\\
${}^2$ St. Petersburg Department of Steklov Mathematical Institute, Russian Academy of Sciences,
Fontanka 27, St. Petersburg, Russia\\
${}^3$ Centre de Physique Theorique de Luminy,
Case 907, Luminy, F-13288 Marseille, France\\
}

\begin{abstract} \noindent{\bf Abstract\ } {\it The Spin Foams for People Without the 3d/ 4d Imagination} could be an alternative title of our work. 
We derive spin foams from {\it operator spin network diagrams}
we introduce. Our diagrams are the spin network  analogy of the Feynman diagrams.
Their framework is compatible with the framework of Loop Quantum Gravity. 
For every operator spin network diagram we construct a corresponding operator spin foam. Admitting all the spin networks of LQG and all possible diagrams leads to a clearly defined large class of operator spin foams. In this way our framework provides a proposal for a class of 2-cell complexes that should be used in the spin foam theories of LQG.  
Within this class, our diagrams are just equivalent to the spin foams.
The advantage, however, in the diagram framework is, that it is self contained, all the amplitudes can be calculated directly from the diagrams without explicit visualization of the corresponding spin foams. The spin network diagram operators and amplitudes are consistently defined on their own. Each diagram encodes all the combinatorial information. We  illustrate applications of our diagrams: we introduce a diagram definition of Rovelli's surface amplitudes as well as of the canonical transition amplitudes. Importantly, our operator spin network diagrams are defined in a sufficiently general way to accommodate all the versions of the EPRL or the FK model, as well as other possible models. The diagrams are also compatible with the structure of the LQG Hamiltonian operators, what is an additional advantage. Finally, a scheme for a complete definition of a spin foam theory by declaring a set of interaction vertices emerges from  the examples presented at the end of the paper. 
 
\end{abstract}
\pacs{
  {04.60.Pp},
  {04.60.Gw}
  }
\maketitle


\section{Introduction}\label{sc:intro}
\subsection{Motivation}
The idea of  spin foam models is to define histories of the spin networks using 2-cell complexes colored by a given group $G$ representations and by intertwiners or equivalently by operators \cite{RR, Markopoulou, Baezintro, perez,Rovellibook,NP,Operator_SF}. The Engle-Pereira-Rovelli-Livine model \cite{EPRL} and Freidel-Krasnov model \cite{FK} (combined with \cite{SFLQG}) relate the spin foams directly with the spin network states, in particular with the spin network states of LQG \cite{Rovellibook,LQGdiscr,AshLewrev,Marev,Ashtekarbook,Thiemannbook}. What is still needed, is a unique definition  of a class of the 2-cell complexes that are taken into account. In the spin foam literature assumptions consistent with a given framework are formulated, and the complexes are: either simplicial \cite{EPRL,ASYMP1}, or cubular \cite{thiemann} or linear \cite{Baezintro}, or locally linear \cite{SFLQG}, or combinatorially defined \cite{Rovelli_Smerlak}, or some other restrictions on the gluing of the 2-disks were made \cite{Operator_SF}, or the spin foams were derived as the Feynman diagrams from  actions of  Group Field Theory models \cite{FreidelGFT,depietri,Geloun:2010vj,Krajewski:2010yq}. 

On the one hand,  the familiar simplicial 2-cell complexes are not sufficient because they do not apply to all the states of quantum geometry according to LQG. On the other hand, though, the classes of the linear or, respectively, locally linear 2-complexes as general as they are, unnecessarily invoke  auxiliary affine spaces, affine structures which are not compatible with the diffeomorphism invariance of GR. Finally, general CW-complexes \cite{Hatcher} allow a diversity that seems to go beyond the graphs and spin networks. 

\subsection{Our goal - the spin network diagrams} 
In the current paper we derive spin foams from {\it operator spin network diagrams}
we introduce. Our diagrams are the spin network  analogy of the Feynman diagrams.
Their framework is compatible with the framework of LQG. 
For every operator spin network diagram we construct a corresponding operator spin foam. Admitting all the spin networks of LQG and all possible diagrams leads to a clearly defined large class of operator spin foams. In this way our framework provides a proposal for a class of 2-cell complexes used in the spin foam theories.  
Within this class, our diagrams are just equivalent to the spin foams.

The advantage in the diagram framework is, that it is self contained, all the amplitudes can be calculated directly from the diagrams without explicit constructing the corresponding spin foams. Indeed, the spin network diagram operators and amplitudes can be consistently defined on their own. Given a diagram the reconstruction of an operator spin foam itself is not necessary, because the diagram encodes all the information. And it is convenient, because using the diagrams is much simpler than using the spin foams, therefore one may call our framework {\it the spin networks for people
without the 3- and 4-dimensional space imagination}.

We  illustrate 
applications of our diagrams: we introduce a diagram definition of  Rovelli's surface amplitudes as well as the canonical transition amplitudes.       
Importantly, our operator spin network diagrams are defined in a sufficiently general
way to accommodate all the versions of the EPRL or the FK model, as well other possible models. The diagrams are also compatible with the framework used in LQG to define the Hamiltonian operators, what is an additional advantage.

Our paper is organized as follows. 

First, we illustrate our idea on a simple non-trivial example in the next subsection, still in Introduction.  

Next, in \sref{diagrams} we introduce general definitions of  graph diagrams and, respectively,  operator spin network diagrams. For the reader's convenience, we
demonstrate how this framework can be applied in a self-sufficient way, giving rise
to operators and amplitudes calculated without explicit visualization of spin foams.  

On the other hand, we also study in detail the transition from the diagrams to the spin foams. We construct explicitly all the 2-cell complexes corresponding to our diagrams. Each of the 2-complexes is characterized by a diagram which consist of a set of graphs endowed with suitable relation in the set of vertices and links which we name {\it graph diagram}. This is a generalization of  diagrams defined by Frank Hellmann \cite{FrankPhD} for the simplicial triangulations. We  introduce general graph diagrams (\sref{diagrams}) and  present an exact, explicit construction of the corresponding 2-cell complexes  (\sref{construction}). Next, we characterise the resulting 2-cell complexes and discuss their properties (\sref{properties}).  

Operator spin network diagrams are defined as suitably colored graph diagrams (\sref{coloring}).
For each given diagram, the coloring passes  to a coloring of the corresponding 
2-cell complex and makes it an operator spin foam \cite{cEPRL,Operator_SF}
(plus the generalizations we include all the EPRL models).

In \sref{examples} we show farther examples of the diagrams and corresponding spin foams. It is easy to identify the elements of the diagram corresponding to to free propagation of the quantum state (the propagator of a spin network) and to the interaction (nontrivial vertices of the spin foam).

\subsection{Example illustrating our idea\label{sc:IntrExample}}
\begin{figure}[hbt!]
	\centering
	\subfloat[The foam $\kappa$]{\label{fig:1_spinfoam}\includegraphics[width=0.3\textwidth]{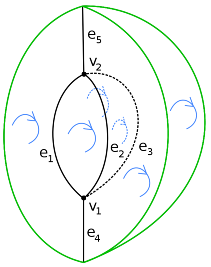} }
	\hspace{0.02\textwidth}
	\subfloat[The foams $\kappa_1$ and $\kappa_2$ corresponding to the vertices $v_1$ and $v_2$.]{\label{fig:twospinfoams}\includegraphics[width=0.3\textwidth]{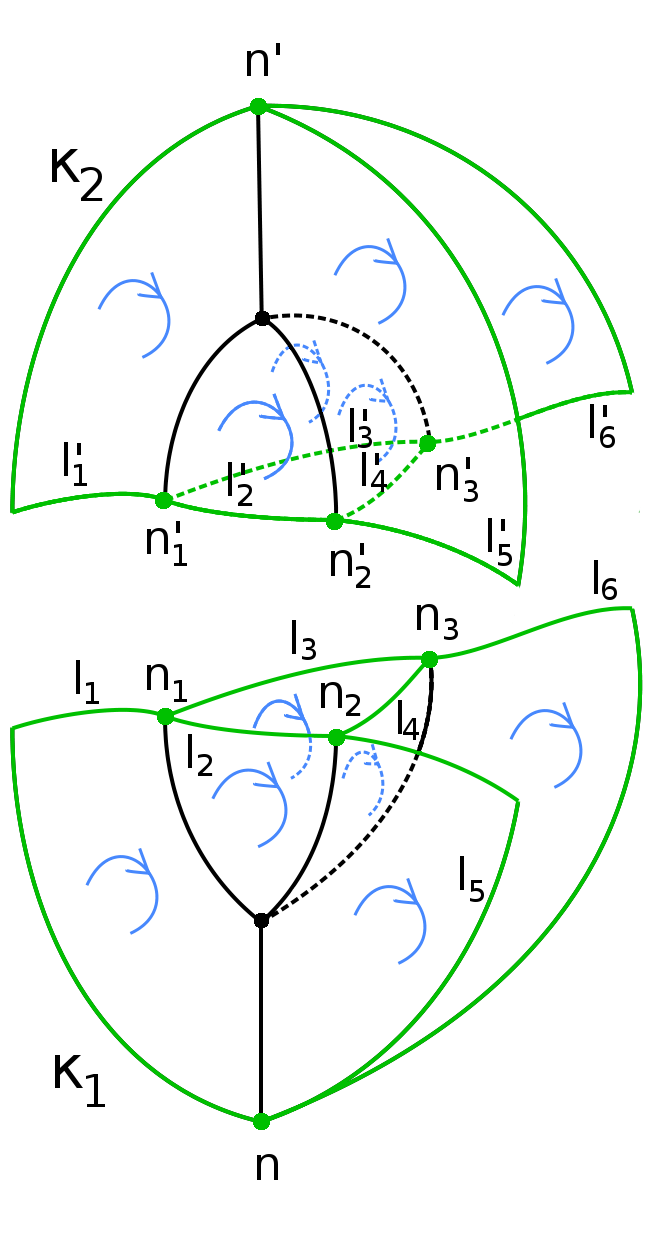}}
	\hspace{0.05\textwidth}
	\subfloat[The graphs corresponding to the foam~$\kappa$.]{\label{fig:spinnetworks}\includegraphics[width=0.3\textwidth]{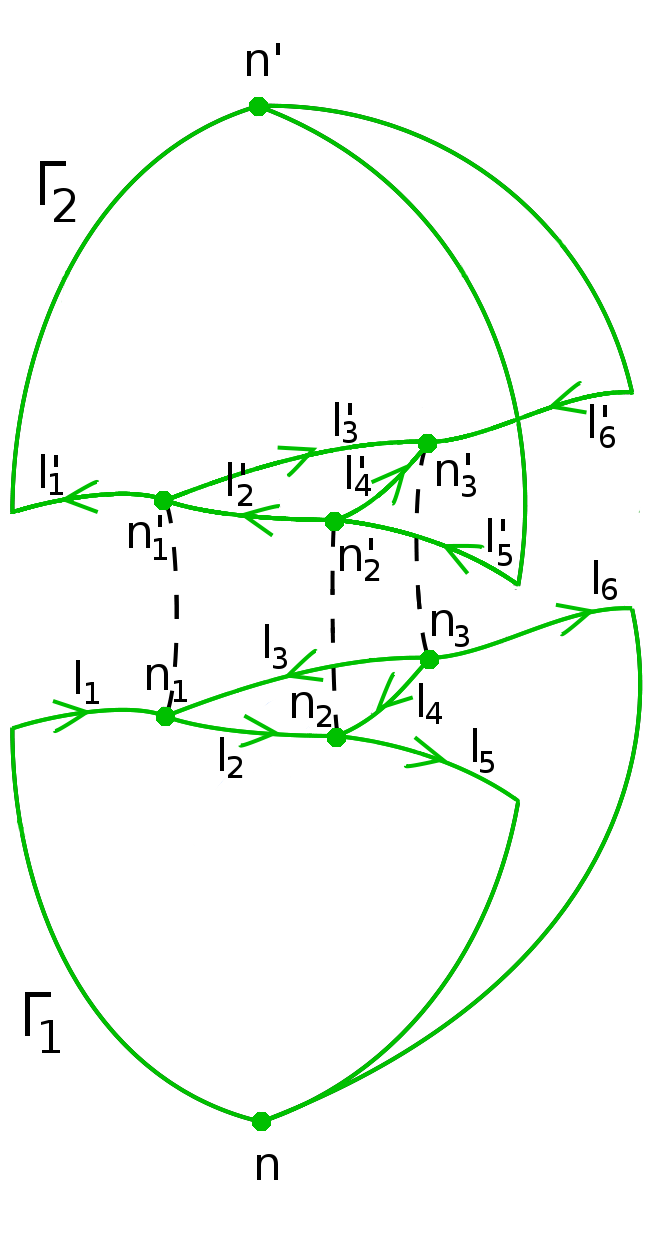}}
	\caption{A example of a foam and the corresponding graphs.}
	\label{fig:1_operator}
\end{figure}
Before the systematic presentation of our construction and results we illustrate our idea on a simple example. We will consider now an operator spin foam defined on a familiar 2-cell complex, and introduce the corresponding diagram.  General definitions of graph diagram and respectively operator spin network diagram will be formulated in the next section. 

Consider a 2-cell complex $\kappa$ depicted on \fref{1_spinfoam} whose boundary is marked by the color green. We will construct the corresponding graph diagram. The 2-cell complex $\kappa$ has two internal vertices $v_1$ and $v_2$. First, cut $\kappa$ into two 2-cell complexes, say $\kappa_1$ and $\kappa_2$ as on \fref{twospinfoams}. The complex $\kappa_1$~($\kappa_2$) is a neighborhood of the vertex $v_1$~($v_2$), and its boundary is the graph $\Gamma_1$~($\Gamma_2$) depicted on the \fref{spinnetworks}. Conversely, given the graph $\Gamma_1$~($\Gamma_2$) on \fref{grafy} (ignore the dashed lines), in order to reconstruct the  2-cell complex $\kappa_1$~($\kappa_2$), one takes a homotopy of $\Gamma_1$~($\Gamma_2$) into a point, denotes the point $v_1$~($v_2$) and views the homotopy as on \fref{twospinfoams}. The image of the homotopy is the complex $\kappa_1$~($\kappa_2$). However, in order to reconstruct the 2-cell complex $\kappa$ one needs the information about gluing of $\kappa_1$ and $\kappa_2$. It is symbolically marked at \fref{grafy} by the dashed curves connecting suitable nodes of the graphs. The complete information can be encoded in the graphs $\Gamma_1$ and $\Gamma_2$ by indicating: $(i)$ the pairs of nodes which should be glued with each other, that is $(n_1,n'_1)$, $(n_2,n'_2)$, $(n_3, n'_3)$, and $(ii)$ the links whose segments are glued at each node, that is at $(n_1,n'_1)$ we glue the segments of  $(l_1, l_2, l_3)$ with the segments of $(l'_1,l'_2,l'_3)$ respectively and similarly at $(n_2,n'_2)$,~$(n_3,n'_3)$. In conclusion, the gluing information can be encoded as a following relation: 
\be\label{eq:relacja}
	\T R\ =\ \{[(n_1; l_1,l_2,l_3), (n'_1;l'_1,l'_2,l'_3)],\ \  [(n_2; l_2,l_5,l_4), (n'_2;l'_2,l'_5,l'_4)],\ \ [(n_3; l_3,l_4,l_6), (n'_3;l'_3,l'_4,l'_6)]\}\ .
\ee
The pair of graphs $\Gamma_1$,~$\Gamma_2$ endowed with the relation $\T R$ is an example of a graph diagram (the general definition below) equivalent to the 2-cell complex $\kappa$.  

\begin{figure}
	\includegraphics[width=0.35\textwidth]{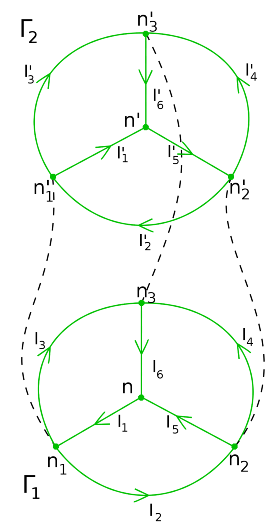}
	\caption{The graph diagram corresponding to $\kappa$. The dashed lines represent the node relation.}
	\label{fig:grafy}
\end{figure}

Next we turn to an example of operator spin network diagram equivalent to an operator spin foam defined using the 2-cell complex $\kappa$.  An operator spin foam can be defined  by a coloring $(\rho,P,A)$ of the elements of the 2-cell complex $\kappa$ (given a Hilbert space $\cal H$, by $\cal H^*$ we denote the dual Hilbert space): $(i)$ $\rho$ is a coloring of its faces with irreducible representations of a given group $G$, $(ii)$ $P$ is a coloring of its non-boundary edges with operators, and $(iii)$ $A$ is a coloring of its internal vertices with contractors, that is tensors used to contract the operators assigned to the edges meeting at the vertex (
the contractor is a new element, we introduce, in order to generalize the notion of operator spin foams \cite{Operator_SF} such that the EPRL model \cite{EPRL} (both euclidean and lorentzian) can be viewed as SU(2) operator spin foam model with the boundary Hilbert space consistent with the LQG kinematical Hilbert space).
\begin{figure}[hbt!]
	\centering
	\subfloat[The operator spin foam $(\kappa,\rho,P)$]{\label{fig:coloredspinfoam}\includegraphics[width=0.40\textwidth]{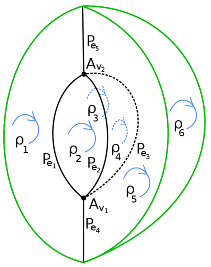} }
	\subfloat[The operator spin network diagram. The dashed lines represent the relation on nodes. $P_{n}=P_{e_4}$, $P_{\{n_1,n'_1\}}=P_{e_1}$, $A_1=A_{v_1}$, etc.]{\label{fig:coloredspinnetworks}\includegraphics[width=0.40\textwidth]{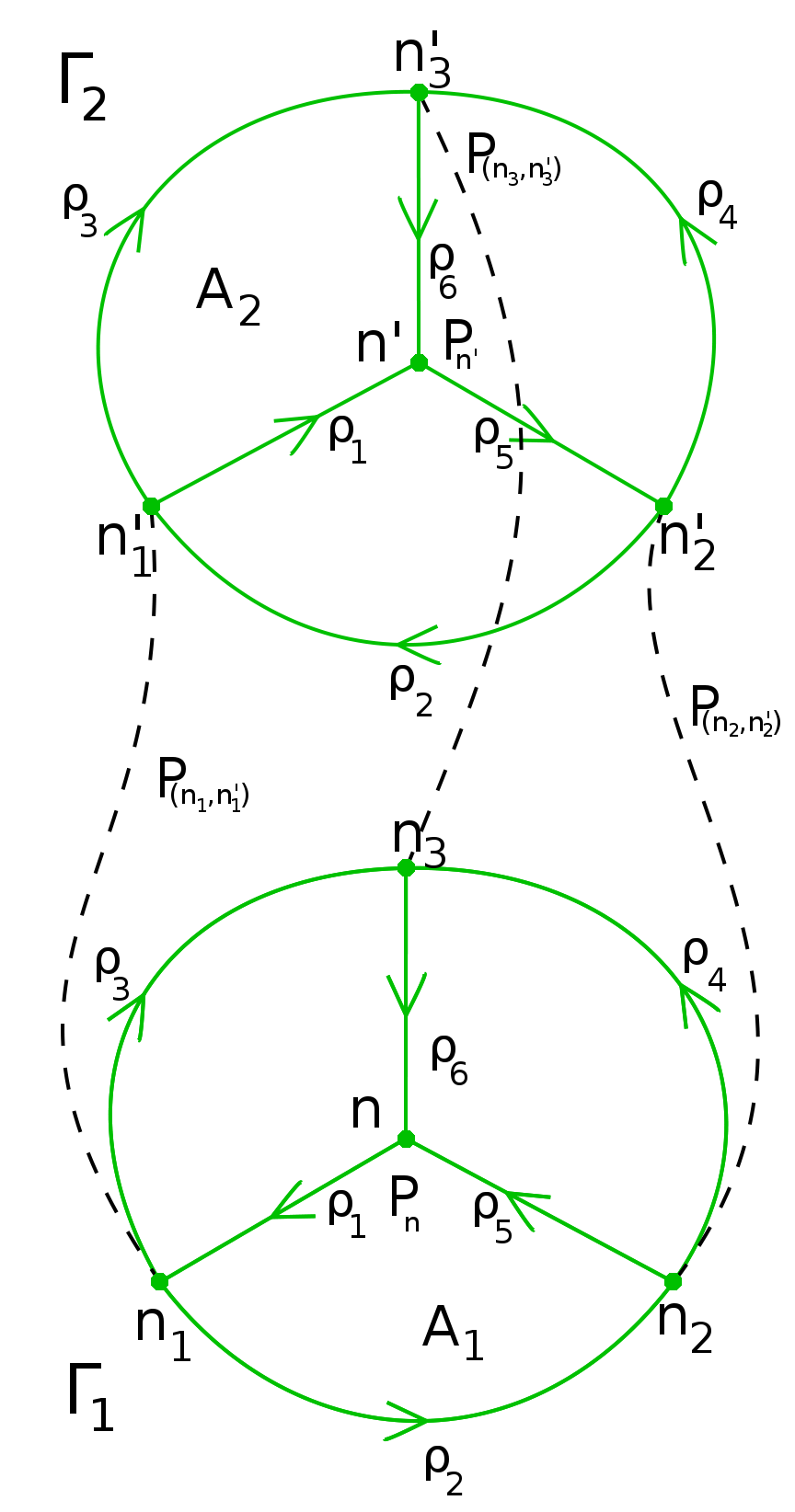}}
	\caption{The operator spin foam and the corresponding operator spin network diagram.}
	\label{fig:11_operator}
\end{figure}

To be specific: each (oriented) face $f_i$ of $\kappa$ is colored by a representation $\rho_i$ of $G$ in the Hilbert space ${\cal H}_i$ (see \fref{1_operator} -- the faces are topological polygons, whose orientations are marked by blue semi-circles). To each  internal edge $e$ (unoriented), i.e. to each one of $e_1$,$e_2$,$e_3$,$e_4$,$e_5$ (see \fref{1_spinfoam}), and each of its end points $w$ , we assign a Hilbert space ${\cal H}_{w,e}$ defined as follows. A face containing the edge $e$ induces an orientation of $e$. According to that orientation the point $v$ is either the beginning or the end of $e$. The definition of the  Hilbert space reads
\be
	\Hil_{w,e}\ =\ \inv\left(\bigotimes_i \Hil_{\rho_i}\otimes \bigotimes_j\Hil^*_{\rho_j}\right)\;\subset\;\bigotimes_i \Hil_{\rho_i}\otimes \bigotimes_j\Hil^*_{\rho_j}
\ee
where $i$~($j$) label the faces such that $v$ is the end~(beginning) of $e$ and Inv stands for subspace of $G$-invariants. Elements of $\inv\left(\bigotimes_i \Hil_{\rho_i}\otimes \bigotimes_j\Hil^*_{\rho_j}\right)\;$ inherit the index structure of the bigger space $\bigotimes_i \Hil_{\rho_i}\otimes \bigotimes_j\Hil^*_{\rho_j}$. For example, given three representations of SU(2): $\rho_1$ and $\rho_2$ of the spin $\frac{1}{2}$ and $\rho_3$ of the spin $1$, each element of $\inv\left(\Hil_{\rho_{1}}\otimes \Hil_{\rho_2}\otimes\Hil_{\rho_3} \right)\;$ is of the form $a\,\tau^{ABi}$, where $a\in\mathbb{C}$ and $\tau$ is the invariant tensor obtained from the Pauli matrices. To the edge $e$ itself we assign the Hilbert space
\be
	{\cal H}_e\ =\ \bigotimes_{w}{\cal H}_{e,w}\ .
\ee
where $w$ runs through the set of the end points of $e$. Notice, that $\Hil_{e,w_1}=\Hil_{e,w_2}^*$, when $w_1$ and $w_2$ are different end points of the edge $e$, so one may interpret every element of ${\cal H}_e$ as an operator $\Hil_{e,w_1}\rightarrow \Hil_{e,w_1}$ or $\Hil_{e,w_2}\rightarrow \Hil_{e,w_2}$. Finally, the edge $e$ is colored by an operator
\be
	 P_e\in {\cal H}_e.
\ee
In this way, the operator coloring $e\mapsto P_e$ is defined for all internal edges of $\kappa$.

The structure of the colorings $\rho$ and $P$ admits at each internal vertex $v$ of $\kappa$ a unique contraction of the operators $P_e$ coloring the edges intersecting $v$. Indeed, to each internal vertex $v$ we assign the Hilbert space 
\be\label{eq:Hil.v}
	{\cal H}_v\ =\ \bigotimes_e {\cal H}_{e,v}
\ee 
where $e$ ranges the set of edges meeting at $v$. In other words:
\be
	{\cal H}_v\ \equiv\ \bigotimes_i {\cal H}_{\rho_i} \otimes {\cal H}_{\rho_i}^*,
\ee 
where $i$ labels the faces of $\kappa$ intersecting $v$.
The natural contraction 
\be
	A^{\rm{Tr}}: {\cal H}_v\rightarrow \mathbb{C},
\ee
is the tensor product of the natural contractions  
\be
	\rm{Tr}_i\ :\ {\cal H}_{\rho_i}\otimes{\cal H}_{\rho_i}^*\rightarrow \mathbb{C}.
\ee
However, to accommodate the EPRL vertex amplitude defined on  an SU(2) spin foam, it is necessary to introduce at the internal vertices  general contractors.
Therefore, for the sake of  both: the naturality and the relevance, we admit all possible contractors, that is we color the internal vertices $v$ by arbitrary elements
\be
	 A_{v}\ \in \ {\cal H}_{v}^*.
\ee 
After decomposing $A_v$ in intertwiner basis, the coefficients becomes so called vertex amplitudes.

Given that operator spin foam defined on the 2-cell complex $\kappa$ by any of the colorings $(\rho, P , A)$, there is a natural contraction   
\be\label{eq:presfoperator}
	\left(A_{v_1}\otimes A_{v_2}\right)\lrcorner\left(P_{e_1}\otimes...\otimes P_{e_5}\right)\ \in\ {\cal H}_{e_4,n}\otimes{\cal H}_{e_5,n'}
\ee
obtained by applying the contractor $A_{v_1}$ to the operators $P_{e_1}$,$P_{e_2}$,$P_{e_3}$,$P_{e_4}$ at $v_1$ and the contractor $A_{v_2}$ to operators $P_{e_1}$,$P_{e_2}$,$P_{e_3}$,$P_{e_5}$ at $v_2$. Here $n$ and $n'$ are the boundary vertices of $\kappa$, the ends of the internal edges $e_4$, and respectively, $e_5$ depicted on \fref{coloredspinnetworks}.   

A spin foam operator $P(\kappa,\rho,P,A)$ we eventually assign to the operator spin foam $(\kappa,\rho,P,A)$ usually involves an extra factor: the product of so called face amplitudes, i.e. numbers $A_f$ assigned to the faces $f$ (typically the dimension of the corresponding representation), and so called boundary edge amplitudes, i.e. numbers $A_e$ assigned to boundary edges $e$ (typically square root of inverse of face amplitude): 
\be\label{eq:sfoperator}
	 P(\kappa,\rho,P,A)\ =\ \left(\prod_{e}A_e \prod_{f}A_f\right)\left(A_{v_1}\otimes A_{v_2}\right)\lrcorner\left(P_{e_1}\otimes\cdots \otimes P_{e_5}\right)
\ee

Let us relate now the current operator spin foam definition with the natural operator spin foam models defined in \cite{Operator_SF} and the EPRL model \cite{EPRL}.

If the operator spin foam introduced above comes from a natural operator spin foam model, then $G$ is a compact group, each of the operators $P_{e_1}, ..., P_{e_5}$ (viewed as a map ${\cal H}_n\rightarrow {\cal H}_n$) is a projection, and each of the contractors $A_{v_1}, A_{v_2}$ is the natural contractor $A^{\rm Tr}$.  

On the other hand if the operator spin foam introduced above comes from the EPRL model,  then $G=$ SU(2), each of the operators $P_{e_1}, ..., P_{e_5}$ (viewed as a map ${\cal H}_n\rightarrow {\cal H}_n$) is the identity map, and each of the contractors $A_{v_1}, A_{v_2}$ is the EPRL contractor $A^{\rm EPRL}$ directly given by the EPRL fusion map \cite{EPRL,SFLQG,Kamykfinite}.                   

Now, after introducing the operator spin foam $(\kappa,\rho,P,A)$ we are in the position to define an equivalent operator spin network diagram. The idea is to encode in the graph diagram $(\Gamma_1,\Gamma_2, \T R)$ (see \fref{grafy} and Eq. \reef{relacja}) the data defining the operator spin foam $(\kappa,\rho,P,A)$: $(i)$ each link $l_1, \ldots, l_6$ ($l'_1, \ldots, l'_6$) of the graph $\Gamma_1$~($\Gamma_2$) \fref{grafy} corresponds to exactly one face of $\kappa$ \fref{1_spinfoam} and inherits its orientation and the representation color $\rho_1, \ldots, \rho_6$ ($\rho_1, \ldots, \rho_6$) - see \fref{11_operator}, $(ii)$  every node $n_1,n_2,n_3,n$ ($n'_1,n'_2,n'_3,n'$) of $\Gamma_1$~($\Gamma_2$) \fref{grafy} corresponds to exactly one internal edge $e_1$,$e_2$,$e_3$,$e_4$ ($e_1$,$e_2$,$e_3$,$e_5$) of $\kappa$ \fref{1_spinfoam}, and inherits the operator $P_{e_1}$,$P_{e_2}$,$P_{e_3}$,$P_{e_4}$ ($P_{e_1}$,$P_{e_2}$,$P_{e_3}$,$P_{e_5}$) \fref{coloredspinfoam} $(iii)$ the graph $\Gamma_1$~($\Gamma_2$) \fref{grafy} itself corresponds uniquely to the internal vertex $v_1$~($v_2$) \fref{1_spinfoam} of $\kappa$ and inherits the contractor $A_1$~($A_2$) \fref{coloredspinfoam}. Every pair of nodes in relation $\T R$ Eq. \reef{relacja} is colored by a same operator, every pair of links in relation $\T R$ is colored by a same representation. 

The resulting operator spin network diagram equivalent to the operator spin foam \fref{coloredspinfoam} is depicted on \fref{coloredspinnetworks} and the information it contains is completed by the relation $\T R$. The spin foam operator $P(\kappa,\rho,P,A)$  is determined by the diagram itself, by taking one operator per each pair of the nodes $(n_i,n_i')$, $i=1,2,3$, being in the relation $\T R$, one operator per each free node $n$ and $n'$, multiplying all of them tensorialy and contracting with the contractors $A_1$ and $A_2$. This defines the operator \reef{presfoperator}. The factor $\prod_{e}A_e \prod_{f}A_f$ present in \reef{sfoperator} involves reconstruction of the faces and the boundary edges of $\kappa$. Since $\kappa$ can be reconstructed from
the diagram, so can be the face and the boundary amplitudes. However, the visualization of the 2-cell complex $\kappa$ is not necessary, and the face and edge amplitudes may be  read  directly from the diagram. That observation motivates the framework we introduce in the next section.



\section{Graph diagrams and operator spin network diagrams}\label{sc:diagrams}
\subsection{Graph diagrams}

\begin{figure}[hbt!]
	\includegraphics[width=0.6\textwidth]{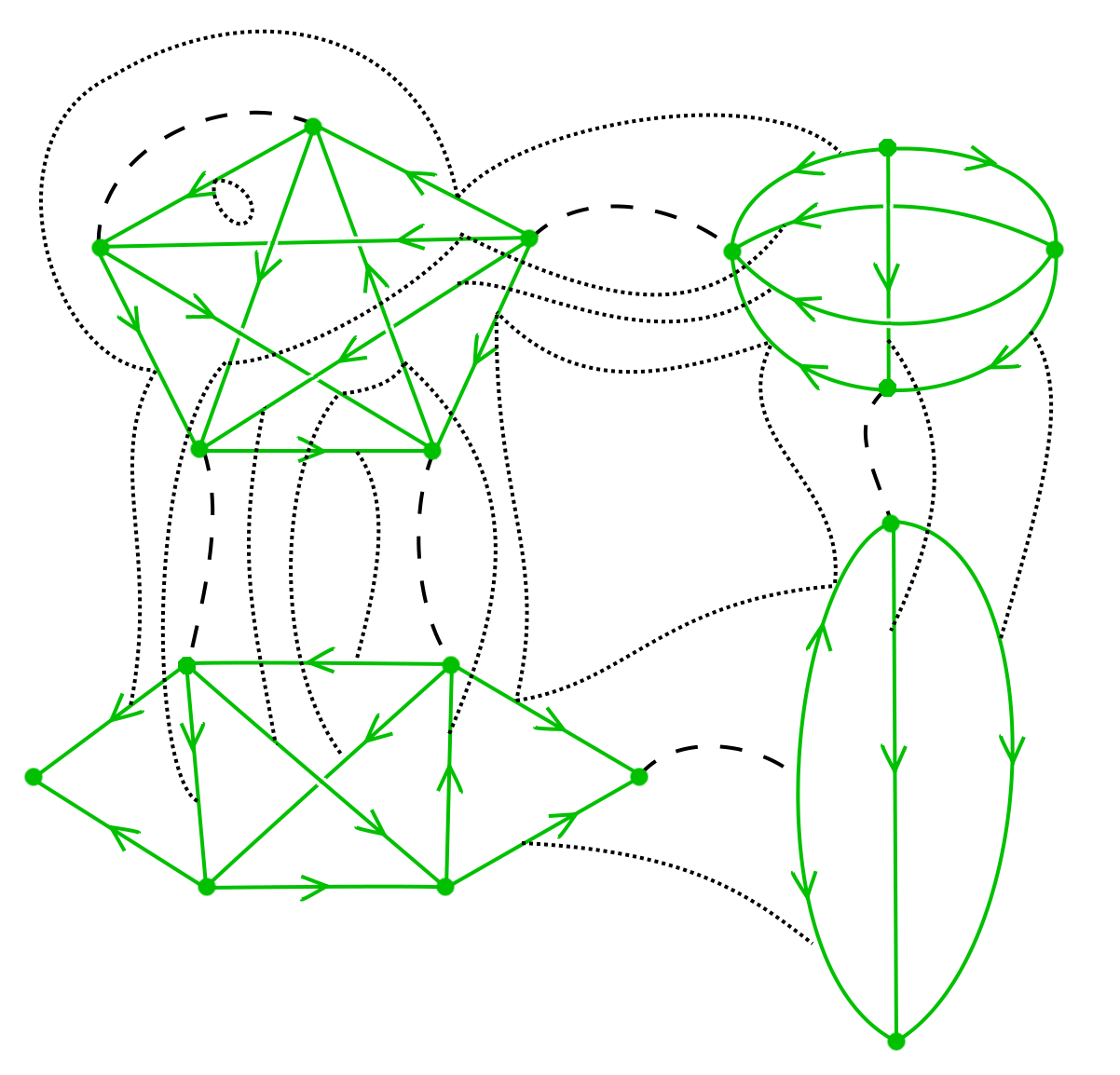}
	\caption{A graph diagram. The thick dots represent the nodes of the graphs, the solid lines with arrows represent the oriented links of the graphs, the dashed lines illustrate the node relation $\T R_{\rm node}$ and the dotted lines illustrate the link relation $\T R_{\rm link}$}
	\label{fig:grafy_relacja}
\end{figure}

A general graph diagram $(\T G,\T R)$ consists of a set ${\T G}$ of oriented graphs $\{\Gamma_1, ... ,\Gamma_N\}$, and a family $\T R$ of relations defined as follows:

\begin{itemize}
	\item $\T R_{\rm node}$: a symmetric relation in the set of nodes of the graphs which we call the node relation, such that 
	each node $n$ either is in relation with precisely one $n'\not=n$ or it is unrelated (and then it is called boundary node).

	\item $\T R_{\rm link}$: a family of symmetric relations in the set of links of the graphs which we call collectively the link relation. If a node $n$ of a graph $\Gamma_I$ is in relation with a node $n'$ of a graph $\Gamma_{I'}$, then one defines a bijective map between incoming~/~outgoing links of $\Gamma_I$ at $n$, with  outgoing~/~incoming links of $\Gamma_{I'}$ at $n'$; no link is left free neither at the node $n$ nor at $n'$; two links identified with each other by the bijection are called to be in the relation $\T R^{(n,n')}_{\rm link}$ at the pair of nodes $n,n'$; a link of $\Gamma_I$~/~$\Gamma_{I'}$ which intersects $n$~/~$n'$ twice,  emerges in the relation twice: once as incoming and once as outgoing.
\end{itemize}

In order to be related , two nodes have to satisfy the consistency condition, that is the number of the incoming~/~outgoing links in each of them has to coincide with the number of the outgoing~/~incoming links at the other one (with possible closed links counted twice). Since two graphs are a graph,
to reduce that  ambiguity we will be assuming that the graphs defining the diagram are connected.  

\subsection{Operator spin network diagrams}
An operator spin network diagram $({\cal G}=\{\Gamma_1,...,\Gamma_N\}, \T R, \rho, P, A)$ is defined by coloring a graph diagram $({\cal G},\T R)$ as follows: 
\begin{itemize}
	\item The coloring $\rho$ assigns to each link $\ell$ of each graph $\Gamma_I$, $I=1,...,N$ an irreducible representation of the group $G$: \be \ell\mapsto\rho_\ell. \ee It is assumed that whenever two links $\ell$ and $\ell'$ are mapped to each other by $\T R_{\rm link}$, then \be\label{eq:RhoSaRowne} \rho_\ell= \rho_{\ell'}.\ee
	\item The coloring $P$ assigns to each node $n$ an operator: \be\label{eq:nodeoperator} n\mapsto P_n\in \Hil_{n}\otimes \Hil_{n}^*, \ee where $\Hil_n$ is defined at each node in the following way:
\be\label{eq:H.n}
	\Hil_n\ =\ \inv\left(\bigotimes_i \Hil_{\rho_i}^* \otimes \bigotimes_j\Hil_{\rho_j}\right)\;\subset\;\left(\bigotimes_i \Hil_{\rho_i}^* \otimes \bigotimes_j\Hil_{\rho_j}\right)
\ee
where $i$~/~$j$ labels the links incoming~/~outgoing at $n$.

 Whenever two nodes $n$ and $n'$ are related by $\T R_{\rm node}$, then (from \reef{RhoSaRowne} and \reef{H.n}) it follows that $\Hil_n=\Hil_{n'}^*$ and  it is assumed about $P$ that \be\label{eq:DualityOfPs} P_n=P_{n'}^*\ee
	\item The coloring $A$ assigns to each graph $\Gamma_I$ a tensor
		\be\label{eq:A.Gamma}
			\Gamma_I\mapsto A_\Gamma\in\ \left(\bigotimes_n {\cal H}_n\right)^*
		\ee
		which we call contractor, where $n$ runs through the nodes of $\Gamma_I$.
\end{itemize}

It is important, that even while saying that $\Hil_n$ consists only of the $G$-invariant elements of  $\left(\bigotimes_i \Hil_{\rho_i}\otimes \bigotimes_j\Hil^*_{\rho_j}\right)$, we think of its elements as being tensor products, elements of the big Hilbert space, possessing the index structure of $\left(\bigotimes_i \Hil_{\rho_i}\otimes \bigotimes_j\Hil^*_{\rho_j}\right)$.

If a node $n$ of one of the graphs $\Gamma_I$ is related by $\T R_{\rm node}$ with another node $n'$ (of the same or different graph), then $P_n$ and $P_{n'}$ are elements of the same Hilbert space $\bigotimes_{n\in\{n,n'\}}{\cal H}_n$; due to \reef{DualityOfPs} they appear to be the same element 
\be\label{eq:P.nn}
	  P_{\{n,n'\}}\ \in \bigotimes_{\tilde n\in\{n,n'\}}{\cal H}_{\tilde n}
\ee 

A natural example of a contractor exists due to the fact that the Hilbert space $\bigotimes_n {\cal H}_n$ 
can be uniquely embedded into with a space:
\be
	\bigotimes_n {\cal H}_n\ \hookrightarrow \bigotimes_i\Hil_{\rho_i}\otimes\Hil_{\rho_i}^*
\ee 
where $i$ ranges the set of the links of $\Gamma_I$. Therefore the distinguished element of $\left(\bigotimes_n {\cal H}_n\right)^*$ is
\be
	 A^{\rm{Tr}}\ =\ \bigotimes_i \rm{Tr}_i
\ee 
which is used e.g. in the $G$-BF theory. This contractor can be also used to define a version of the Euclidean EPRL model (viewed from the $Spin(4)$ spin network perspective \cite{SFLQG,cEPRL,Operator_SF}). However the $SU(2)$ spin foam model constructed from the EPRL vertex amplitude defines a different contractor, that can be denoted by $A^{\rm EPRL}$ \cite{EPRL_contractor, EPRL_full}.       

\subsection{The spin network diagram operator\label{sc:theSFoperator}}
Now we would like to define, given an operator spin network diagram $(\T G,\T R; \rho,P,A)$, an operator analogous to \reef{sfoperator}.

There is a canonical contraction 
\be\label{eq:contraction}
	\tilde P= \left(\bigotimes_I A_I \right)\lrcorner\left(\bigotimes_nP_n\right)
\ee
where $I$ ranges the set $\T G$ of the graphs, and $n$ ranges the set of boundary nodes and the set of pairs of related nodes (with respect to $R_{\rm node}$). 
It is defined by contracting each $A_I$ with the ${\cal H}_n$-part of each operator $P_n\in\Hil_n\otimes\Hil_n^*$ assigned to a boundary node $n$ and the ${\cal H}_{n}$ part of each operator $P_{\{n,n'\}}\in\bigotimes_{\tilde n\in\{n,n'\}}\Hil_{\tilde n}$ assigned a node $n$ related to $n'$, where $n$ ranges the nodes of $\Gamma_I$. 
In the consequence, for each boundary node $n$ one index of $P_n$ remains uncontracted, thus $\tilde P$ is an element of
\be
	\tilde P\in \bigotimes_{{\rm boundary}\; n}\Hil_n^*
\ee

A comparison with the \reef{sfoperator} shows that we are still missing the face and the boundary amplitudes. Therefore an operator defined by operator spin network diagram should have the following form:
\be\label{eq:doperator}
	 P\ =\ \left( A_{\rm boundary} A_{\rm face} \right)\left(\bigotimes_I A_I\right)\lrcorner\left(\bigotimes_nP_n\right),
\ee
where  boundary and face amplitudes $A_{\rm boundary}$ and $A_{\rm face}$ have to be defined suitably as well as the boundary edges and faces themselves. 

The boundary amplitude is given by the product over all the boundary nodes $n$ of graph diagram of amplitudes assigned to the links $\ell$ intersecting these nodes.
\be
	A_{\rm boundary}=\sqrt{\prod_{{\rm boundary}\; n}\prod_{\ell}A_\ell}
\ee
The square root comes from the fact, that each link is counted twice (once per each end).

The face amplitude is given by the product over equivalence classes $f$ of the face relation $\T R_{\rm face}$ (which will be introduced in the subsequent subsection) of the face amplitude
\be
	A_{\rm face}=\prod_{f}A_f
\ee

To specify the numbers $A_\ell$ and $A_f$ we need two functions defined on the space of irreducible representations of $G$:
\be
	A_\ell=f_1(\rho_\ell)\taab A_f=f_2(\rho_f)
\ee
where $\rho_f=\rho_{\ell'}$ for $\ell'$ being any representative of the equivalence class $f$ (we shall see below, that the labeling $\rho$ is consistent with the face relation).

\subsection{Face and edge relations\label{sc:FaceRelation}}
The node relation $\T R_{\rm{node}}$ and the link relation $\T R_{\rm link}$ introduced with the definition of graph diagram  at the beginning of this section lead to equivalence relation in the set of all links of all graphs. The resulting equivalence relation, which we call face relation and denote $\T R_{\rm face}$, carries information about faces of the corresponding 2-complex, and allows to introduce face amplitude without explicit reference to the complex itself. Given a graph diagram $(\T G, \T R)$ the relation $\T R_{\rm face}$ is defined as determined by the following properties:


\begin{itemize}
	\item $\T R_{\rm face}$ is an equivalence relation in the set of all the links $\T L$ of the graphs belonging to $\T G$ 
	\item Two different links $\ell$ and $\ell'$ are in relation $\T R_{\rm face}$ when they are in the relation $\T R^{(n)}_{\rm link}$ in some node $n$.
\end{itemize}

For the later convenience let us characterise possible equivalence classes of $\T R_{\rm face}$:
\begin{enumerate}
	\item Each link unrelated to any other link by any of the $\T R_{\rm link}^{(n,n')}$ relations sets a one-element equivalence class. We will call it an \emph{open} equivalence class.
	\item Each link $\ell$ related to itself by the link relation $\T R_{\rm link}^{(n,n')}$ (at each of its ends $n=s(\ell),n'=t(\ell)$)  sets a one-element equivalence class. We will call it a \emph{closed} equivalence class.
	\item Each pair of links $(\ell_1,\ell_2)$ such that $\ell_1$ and $\ell_2$ are related by the relation $\T R_{\rm link}^{(n,n')}$ and neither $\ell_1$ nor $\ell_2$ is related by relations from $\T R_{\rm link}$ with any other link, sets a two element equivalence class.

		However there are two subcases of such equivalence classes:
		\begin{enumerate}
			\item If $\ell_1$ and $\ell_2$ are related by $\T R_{\rm link}^{(n,n')}$ only at one pair of their ends $n,n'$, the equivalence class will be called \emph{open} and will refer to an external face.
			\item If $\ell_1$ and $\ell_2$ are related by $\T R_{\rm link}^{(n,n')}$ at both pairs of their ends, the equivalence class will be called \emph{cyclic} and will refer to an internal face.
		\end{enumerate}
	\item Every $k>2$-element sequence of links $(\ell_1,\ell_2,\ldots,\ell_k)$ such that ${\ell_i}$, $\ell_{i+1}$, $i\in\{1,\ldots,k-1\}$ are in relation $\T R^{(n_i,n_i')}_{\rm link}$ and the links $\ell_1,\ell_k$ are not in the relation $\T R_{\rm link}$ with any links not belonging to the sequence (i.e. the sequence is \emph{maximal}) sets a $k$-element equivalence class.
		
		Again there are two subcases of such equivalence classes:
		\begin{enumerate}
			\item If $\ell_1$ and $\ell_k$ are not related by $\T R_{\rm link}^{(n,n')}$ at any pair of nodes, the equivalence class will be called \emph{open} and will refer to an external face.
			\item If $\ell_1$ and $\ell_k$ are related by $\T R_{\rm link}^{(n,n')}$ at some pair of nodes, the equivalence class will be called \emph{cyclic} and will refer to an internal face.
		\end{enumerate}

\end{enumerate}

  Each equivalence class $f$ of $\T R_{\rm face}$ will be called merged face. Notice that given an operator spin network diagram $(\T G, \T R, \rho, P ,A)$ two links which belong to a same merged face are colored by a same representation, what was used in the definition of the face amplitude.

For later convenience we will also introduce the \emph{edge} relation $\T R_{\rm edge}$ being determined by the following two properties:        
\begin{itemize} 
	\item $\T R_{\rm edge}$ is an equivalence relation in the set of nodes $\T N$ of all graphs belonging to $\T G$.
	\item Two different nodes $n$ and $n'$ are in relation $\T R_{\rm edge}$ when they are in the relation $\T R_{\rm node}$.
\end{itemize}
There are two types of the equivalence classes:
\begin{enumerate}
	\item Each node unrelated to any other node by $\T R_{\rm node}$ sets a one-element equivalence class.
	\item Each pair of nodes $\{n,n'\}$ related by the $\T R_{\rm node}$ sets a two-element equivalence class.
\end{enumerate}

\subsection{Boundary graph of operator spin network diagram}
Given operator spin network diagram $(\T G, \T R, \rho, P, A)$, we define now its boundary. We will use a new operation defined  on graphs -- merging graphs -- and naturally extend it to  spin networks.

Given two nodes $n$, $n'$ in the graph diagram $(\T G,\T R)$, related by the relation $R_{\rm node}$ (they may be either nodes of a same graph or of two different graphs) the merging is defined in the following way (\fref{node_merging}) 
\begin{enumerate}
	\item Remove these nodes from the graphs they belong to, together with segments of the links meeting at the nodes $n$, $n'$. From each link we remove a segment containing the node $n$ or $n'$ respectively.

\begin{figure}[hbt!]
	\centering
	\subfloat[A pair of related nodes. The dotted/dashed lines represent the link/node relation]{\label{fig:2_spinfoam}\includegraphics[width=0.3\textwidth]{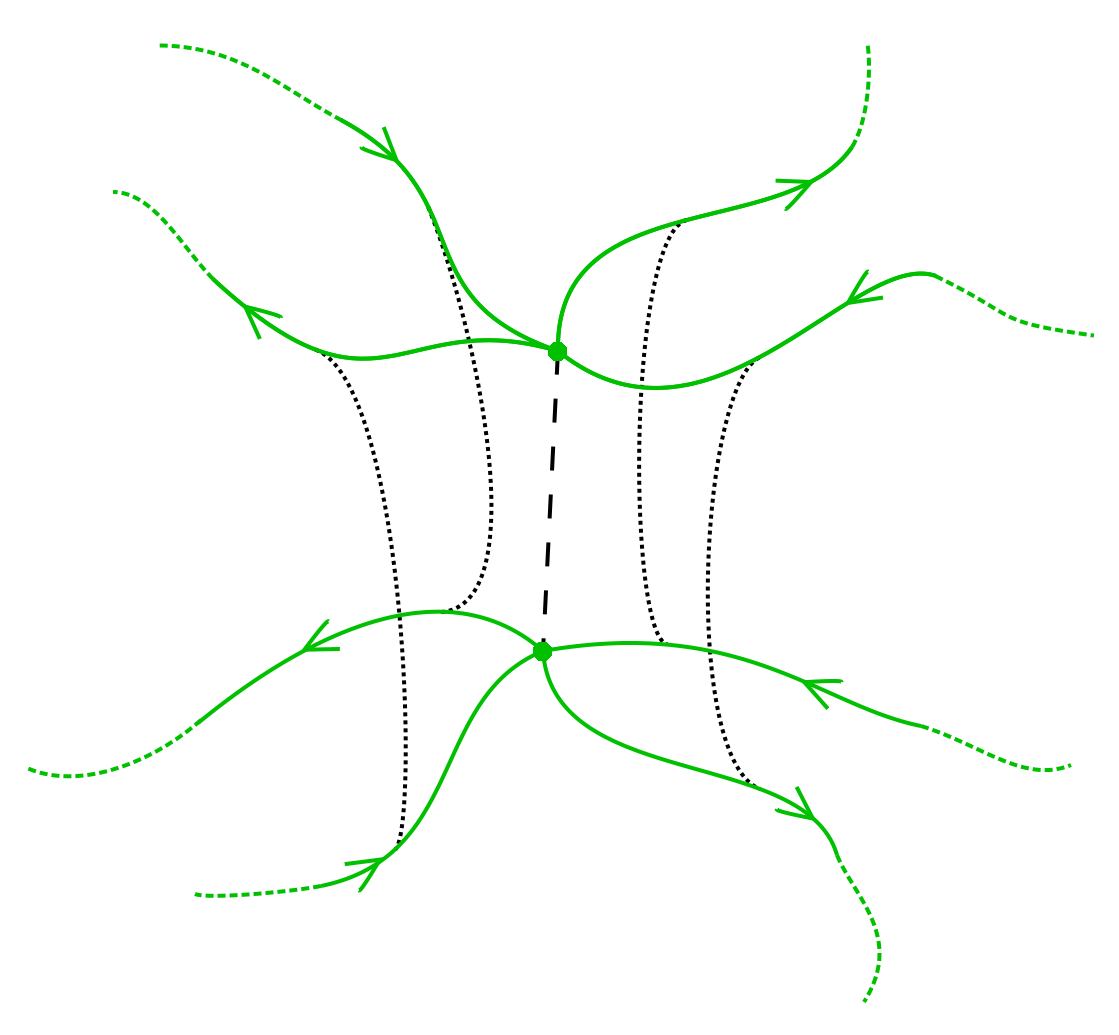} }
	\hspace{0.02\textwidth}
	\subfloat[The neighbourhoods of the nodes are removed]{\includegraphics[width=0.3\textwidth]{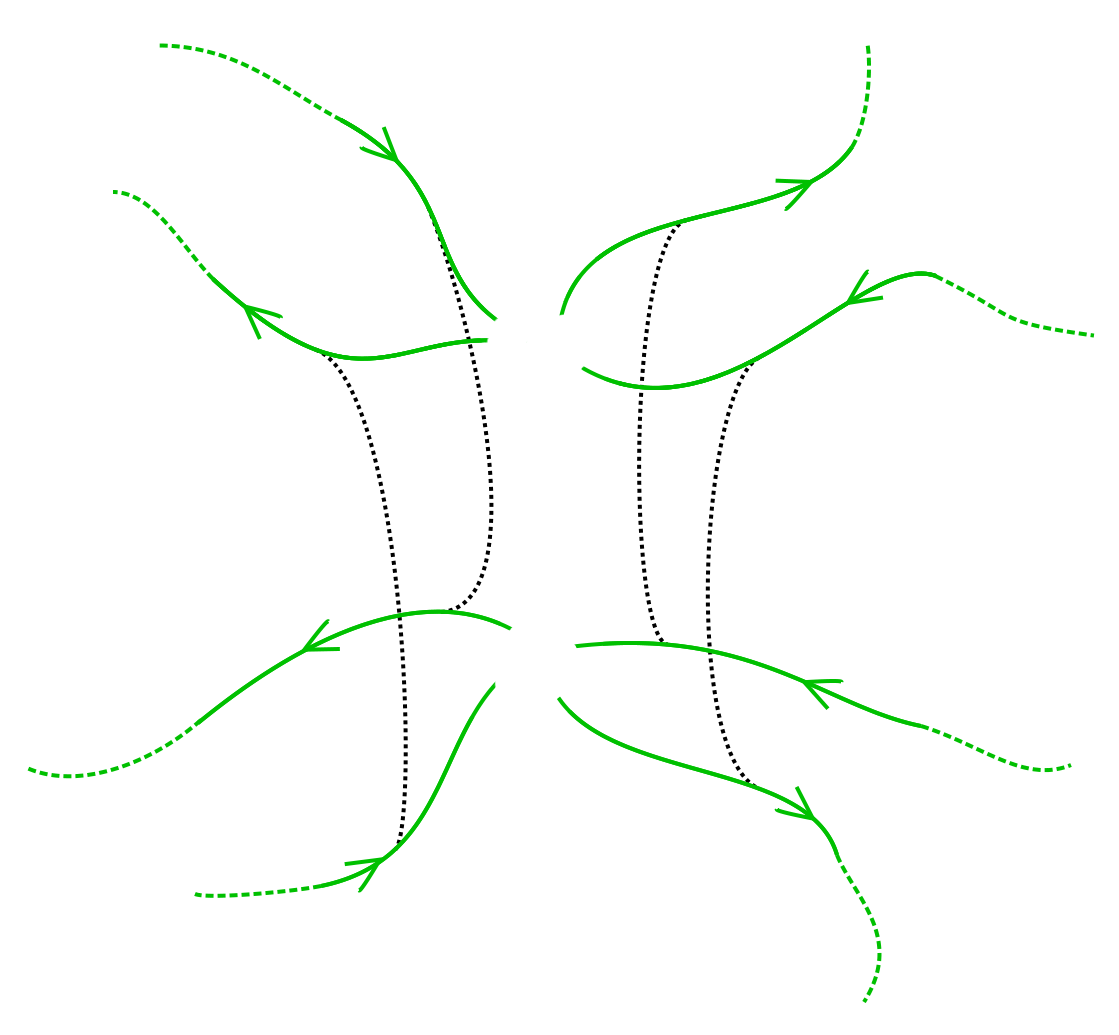}}
	\hspace{0.05\textwidth}
	\subfloat[The related links are connected]{\includegraphics[width=0.3\textwidth]{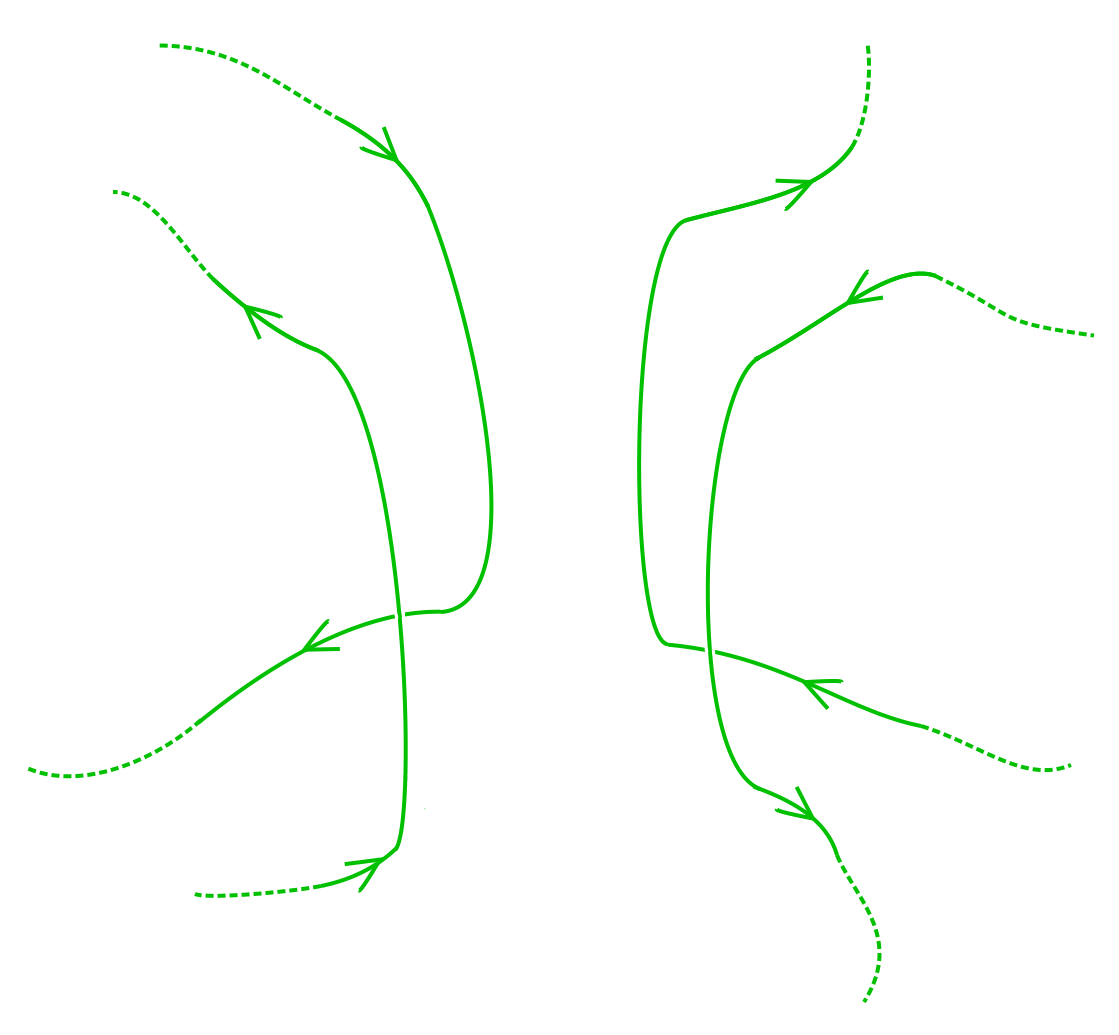}}
	\caption{Merging related nodes in a graph diagram.}
	\label{fig:node_merging}
\end{figure}

	There are two degenerate cases requiring additional instructions:
	\begin{enumerate}
		\item a link is a loop which begins and ends at $n$ (or $n'$). Then we remove both an incoming segment and an outgoing segment.
		\item a link connects the nodes $n$ and $n'$. Then we remove the entire link
	\end{enumerate}
	
	\item We are left with a number of remaining open segments of links. We connect the links that had started/ended at $n$ with the links that had ended/started at $n'$ according to the relation $\T R_{\rm link}$.
\end{enumerate}

\begin{figure}[hbt!]
	\centering
	\subfloat[Two graphs in a graph diagram.]{\label{fig:diagram}\includegraphics[width=0.4\textwidth]{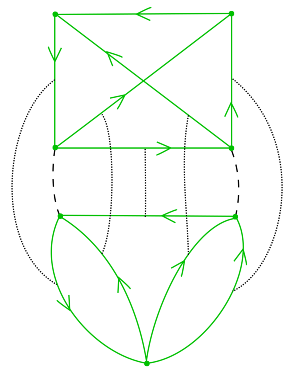} }
	\subfloat[Remove the nodes in the relation.]{\label{fig:remove_nodes}\includegraphics[width=0.4\textwidth]{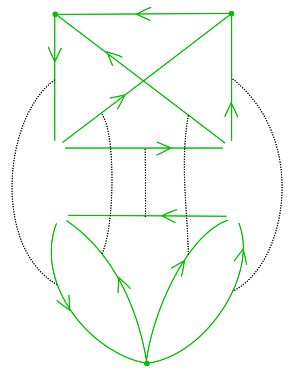}}\\
	\subfloat[Remove each link whose both endpoints where removed.]{\label{fig:remove_links}\includegraphics[width=0.4\textwidth]{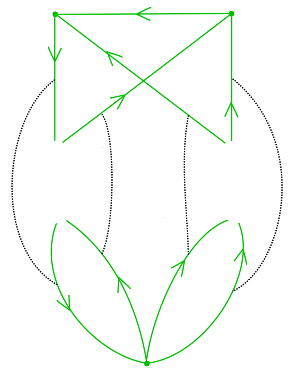} }
	\subfloat[Merge the remaining links.]{\label{fig:merge_links}\includegraphics[width=0.4\textwidth]{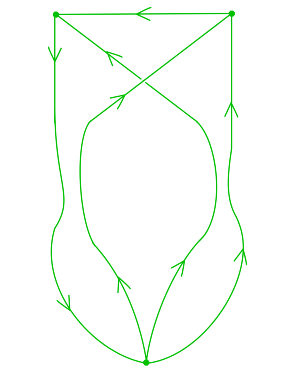} }
	\caption{Merging of graphs in a graph diagram.}
	\label{fig:new_operation}
\end{figure}

The result of merging of the pair of nodes is a new graph diagram  $(\T G',\T R')$. We repeat the merging for another pair of related nodes. We go on until we reach the stage, at which the resulting graph diagram has no pair of related nodes, that is it consists of a set of graphs $\T G_{\rm final}$, with no relation. The graphs constitute the boundary graph (disconnected, if $\T G_{\rm final}$ contains more than one graph).

The resulting boundary graph does not depend on the order in which we merge the pairs of nodes. Proof of that fact, together with more detailed construction, can be found in \sref{construction}. Equivalently, it is easy to perform the merging simultaneously at all the pairs of related nodes. We illustrate it at \fref{brzegowy}.

Coloring $\rho$ of a graph diagram introduced with the definition of operator spin network diagram is consistent with merging of pairs of nodes. Indeed, each pair $\ell$, $\ell'$ of merged links is labeled by same representation $\rho_\ell=\rho_{\ell'}$. Therefore the boundary graph of operator spin network diagram inherit labeling of links by the representations of $G$.

At the beginning of this section we defined boundary nodes of graph diagram. The nodes of the very boundary graph are precisely the boundary nodes defined before. Moreover the links of graph diagram meeting at the boundary nodes (the same that give contribution to the boundary amplitude) are precisely the links of the boundary graph.

\begin{figure}[hbt!]
	\centering
	\subfloat[A graph diagram.]{\includegraphics[width=0.5\textwidth]{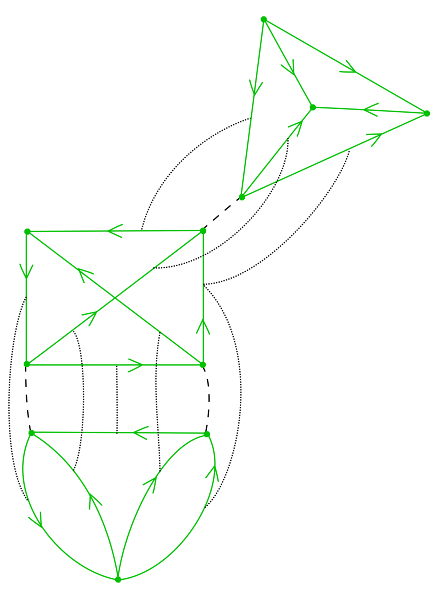} }
	\subfloat[The boundary graph is obtained by merging all the pairs of the related nodes.]{\includegraphics[width=0.5\textwidth]{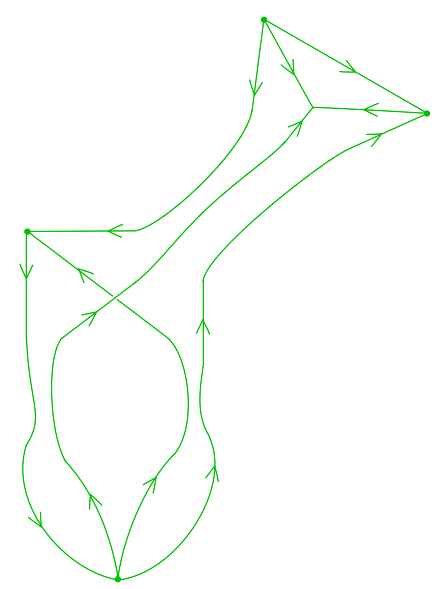} }
	\caption{A graph diagram and the corresponding boundary graph.}
	\label{fig:brzegowy}
\end{figure}

\subsection{Examples of application}
\subsubsection{The EPRL model and the natural operator spin network models\label{sc:ExamplesOfApplication}}
The EPRL model defines the following operator spin network diagrams $(\kappa, \rho, P, A)$. The group is SU(2), the coloring $\rho$ takes values in all the irreducible representations, the values of the coloring $P$ are the identity operators
(viewing each $P_n$ and each $P_{\{n,n'\}}$ as a map ${\cal H}_n\rightarrow{\cal H}_n$), and the coloring $A$ takes  values $A^{\rm EPRL}_v$ given by the fusion map
(either Euclidean or Lorentzian) \cite{EPRL,SFLQG,Kamykfinite}.    
 
Given a natural operator spin network model \cite{Operator_SF}, the group is an arbitrary compact $G$, the coloring $\rho$ takes values in the set of all the irreducible representations,  the coloring $A$ for every internal vertex takes the value $A_v=A^{\rm Tr}$. The coloring $P$ takes values in the projection operators (including the zero operator) which are not specified, however they are constrained by the naturality conditions, the most important are:     each operator $P_n$ and $P_{n,n'}$ is determined by the sequence (unordered, with repetitions) of the representation colors of the links intersecting a given node $n$ (regardless of the structure of the other parts of the diagram), and in the case of a sequence $\rho_1, \rho_1^*$ the projection is not zero (see \cite{Operator_SF} for the details).       

\subsubsection{From the diagrams to Rovelli's  boundary functionals}
Consider a general spin network operator diagram $(\T G,\T R; \rho, P,A)$. Define the Hilbert space of Rovelli's boundary states to be the tensor product labeled by the boundary nodes:
\be
	\Hil_b = \bigotimes_{{\rm boundary}\;n}\Hil_n
\ee
The spin network diagram operator $P$ defined by the formula \reef{doperator} is an element of $\Hil_b^*$.

Given a state $\psi_b\in\Hil_b$ the Rovelli amplitude defined by the diagram is the natural contraction
\be
	\braket{W|\psi_b}:=\psi_b\lrcorner P
\ee

\subsubsection{Rovelli's boundary transition amplitude as a spin network operator diagram}
Given an operator spin network diagram $\T D= (\T G,\T R; \rho, P,A)$ and a boundary state $\psi_b$ (as defined in the previous example) construct an extended operator spin network diagram $\T D'$ defined as:
\begin{itemize}
	\item The new set of graphs is $\T G\sqcup\{\gamma_b\}$, where $\gamma_b$ is the boundary graph of $\T D$ with link orientation reversed.
	\item The new relation $\T R'$ is the relation $\T R$ extended by the pairs $(n_b,n_b')$, where $n_b$ are the boundary nodes of $\T D$ and $n_b'$ are corresponding nodes of the graph $\gamma_b$.
	\item The coloring $\rho$ induces the coloring of the links of $\gamma_b$ (see \reef{RhoSaRowne})
	\item The coloring $P$ induces the coloring of the nodes of $\gamma_b$ (see \reef{DualityOfPs})
	\item The contractor labeling $A$ is extended by $A_{\gamma_b}=\psi$ (the contractor space $\Hil_{\gamma_b}^*=\Hil_b^{**}=\Hil_b$, because links of $\gamma_b$ are reversed links of the boundary graph).
\end{itemize}

The spin network diagram operator $P'$ corresponding to $\T D'$ is a complex number:
\be
	P'=\psi_b\lrcorner P=\braket{W|\psi_b}
\ee

\subsubsection{Simplified formalism - one vertex interaction}

Consider an operator spin network diagram whose set $\T G$ consists of 2 graphs: $\Gamma_\inn$ and $\Gamma_{\rm int}$. The relation $\T R$ does not relate any pair of nodes of $\Gamma_\inn$, otherwise it is arbitrary. Coloring $\rho$ is arbitrary, coloring $P$ is restricted only by the condition that each $P_n$ is a projection. The contractors are arbitrary. However the contractor $A_{\Gamma_\inn}$ is given a special meaning, i.e. it is considered to be the initial state $\psi_\inn\in\Hil_{\Gamma_\inn}$.

The boundary graph of this operator spin network diagram is interpreted as $\Gamma_\out$ (see \fref{final}). The boundary Hilbert space is thought of being the space of final states $\psi_\out$. Given a state $\psi_\inn$ (encoded in the contractor $A_{\Gamma_\inn}$) the spin network diagram operator $P$ is the final state $\psi_\out$ of the interaction described by $A_{\Gamma_{\rm int}}$.

\begin{figure}[hbt!]
	\centering
	\subfloat[An initial graph and an interaction graph]{\label{fig:inital}\includegraphics[width=0.40\textwidth]{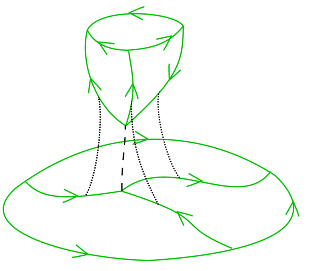}}
	\subfloat[The final graph]{\label{fig:final}\includegraphics[width=0.40\textwidth]{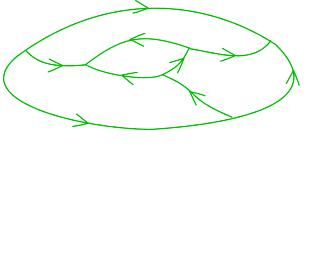}}
	\caption{An example of the 1-vertex interaction: a graph diagram}
	\label{fig:simplified_1}
\end{figure}

\begin{figure}[hbt!]
	\centering
	\subfloat[An initial spin network. In the diagram it becomes a contractor $A_{(\gamma,\rho,\iota)}$]{\label{fig:initial_spinnetwork}\includegraphics[width=0.40\textwidth]{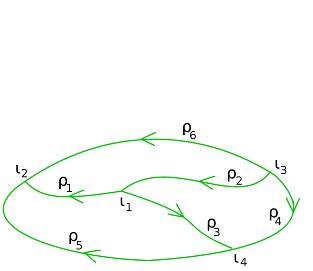}}
	\subfloat[The operator defined by this diagram  is the final spin network.]{\label{fig:inital_interaction_spinnetwork}\includegraphics[width=0.40\textwidth]{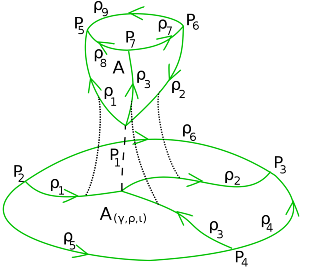}}
	\caption{The example of 1-vertex interaction: the operator spin network diagram.}
	\label{fig:simplified_spinnetwork}
\end{figure}


\section{Construction of a 2-complex defined by a graph diagram\label{sc:construction}}

In the previous section we introduced operator spin network diagrams. The example shown in \sref{intro} explains how to obtain an operator spin network diagram out of an operator spin foam. In the present section we pass from the graph diagrams to 2-complexes. Later (in \sref{coloring}) we will show how the coloring of a graph diagram induce the coloring of the 2-complex what will enable us to construct an operator spin foam out of an arbitrary operator spin network diagram.

It will be convenient to introduce a notion of a \emph{squid graph} being a decomposition of a graph into a set of simpler (open) graphs, called squids (see \sref{squidgraph}). Such decomposition will make the relation $\T R$ (introduced in the previous section) easier to deal with by making it a relation on a set of all squids of all graphs in the graph diagram. Thus our initial data will be a \emph{squid} graph diagram $(\K G,\K R)$ being a set of \emph{squid}-graphs $\K G$ and a set of pairs of squids $\K R$ (together with appropriate maps $\phi_r$ for each $r\in \K R$) such that each squid belongs at most to one pair $r\in \K R$ and whenever a pair of squids belongs to $\K R$ then the two squids are  homeomorphic.

The construction will be as follows: first we will see how to construct an 2-$\Delta$-complex out of an arbitrary squid-graph. Each such complex refers to a spin foam with one vertex. Then we will see, how to glue two such complexes along one pair of squids $r\in \K R$, to obtain a 2-vertex foam. Finally we will see, that the gluing procedure does not require its objects to be 1-vertex foams and it has straightforward generalization to whatever foam, so one can proceed gluing until all the set $\K R$ was used.

\subsection{The squid graph\label{sc:squidgraph}}
We will use the following notation: an oriented graph $\Gamma$ is a pair $(\T N,\T L)$ where $\T N=\{n_1,\ldots,n_N\}$ is a set of nodes, $\T L=\{\ell_1,\ldots,\ell_L\}$ is a set of oriented links. For every link $\ell$ we will denote its beginning (source) by $s(\ell)$ and its end (target) by $t(\ell)$.

\begin{figure}[hbt!]
  \centering
\includegraphics[width=0.40\textwidth]{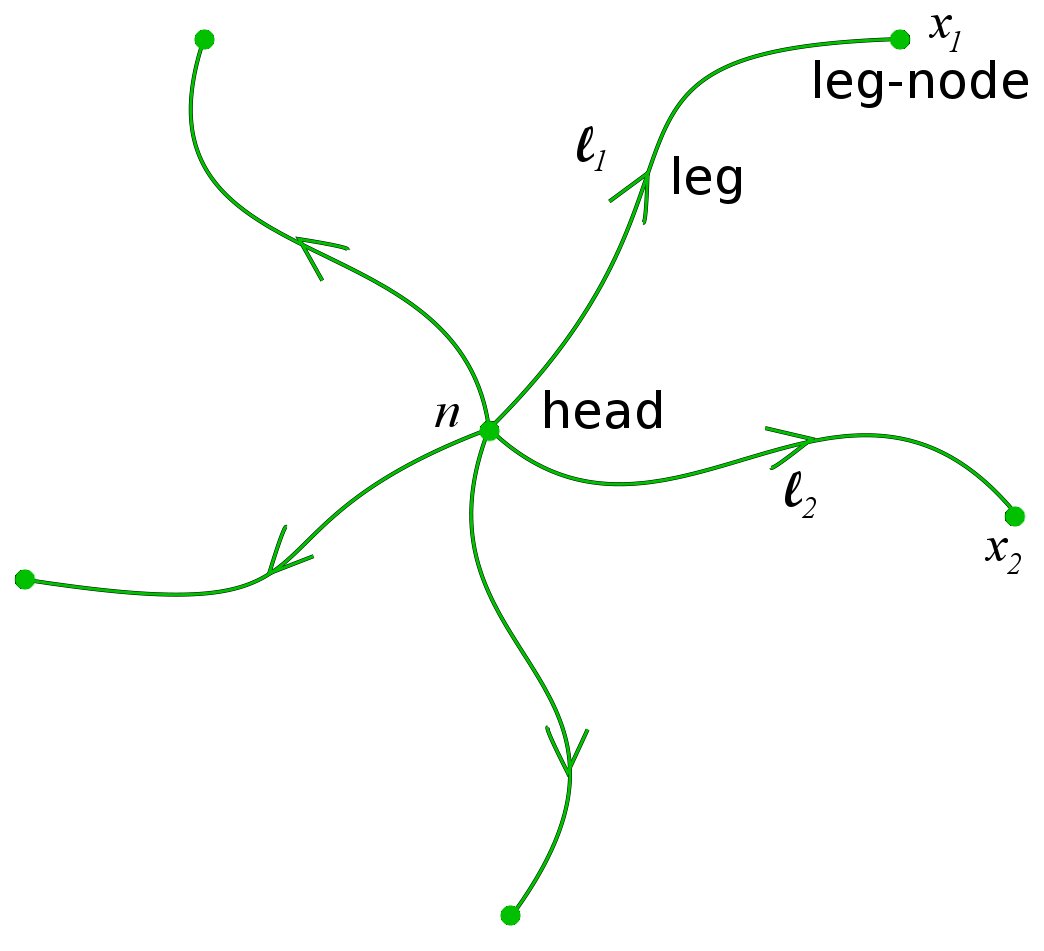}
  \caption{A squid}
  \label{fig:squid}
\end{figure}

\begin{df}
	A \emph{squid} is an oriented graph $\lambda=\left( \{n\}\cup\{x_1,\ldots,x_k\}, \{\ell_1,\ldots,\ell_k\}\right)$ such that each link $\ell_i$ satisfies the condition $s(\ell_i)=n\land t(\ell_i)=x_i$. The node $n$ is called the \emph{head} of the squid. The links $\ell_1,\ldots,\ell_k$ are called the \emph{legs} of the squid and the nodes $x_1,\ldots,x_k$ are called the \emph{leg-nodes} of the squid. (see \fref{squid})
\end{df}
For our applications it is convenient to assume, that the number of legs $k$ is equal or greater than~2. This is because we will glue the squids in order to construct closed graphs from them. 

One can provide either combinatorial definition of a squid graph as a set of squids with a rule of identifying of their boundaries, or a geometrical one: as a decomposition of an ordinary graph into some squids. The later one reads as follows.

Given a graph $\Gamma=(\T N, \T L)$ split each link $\ell_i$ into two links by introducing a new node $x_i$. Then reorient the new links in such a way, that each of them begins at the old node and ends at the new node $x_i$ (see \fref{squidgraph}). The resulting graph is
\be\label{eq:GammaS}
	\s\Gamma = \left(\tilde {\T N} =\T N\cup\{x_\ell\}_{\ell\in \T L}\;;\; \tilde {\T L}\right)
\ee
As a result each old node $n\in\T N$ becomes a head of a squid $\lambda_n$, whose legs are the links of $\s\Gamma$ intersecting $n$  (when not needed, we will drop the subscript $n$).

\begin{figure}[hbt!]
	\centering
	\subfloat[A graph $\Gamma$]{\includegraphics[width=0.30\textwidth]{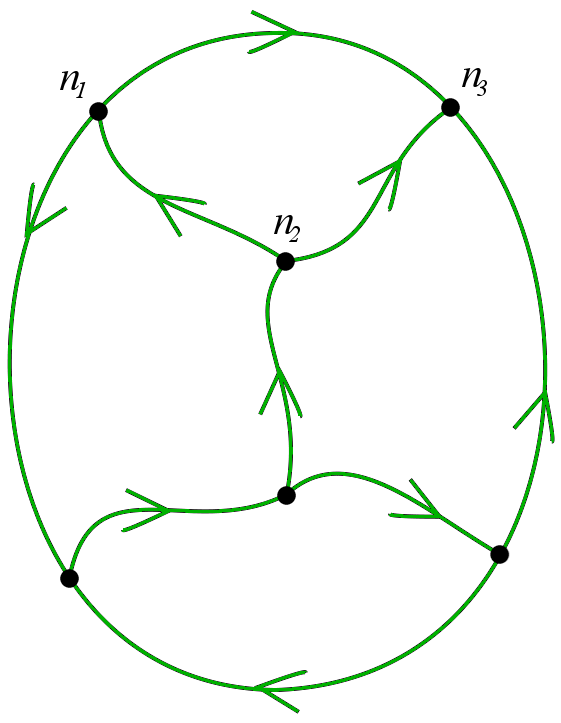} }
	\hspace{0.1\textwidth}
	\subfloat[The corresponding squid graph $\s\Gamma$]{\includegraphics[width=0.30\textwidth]{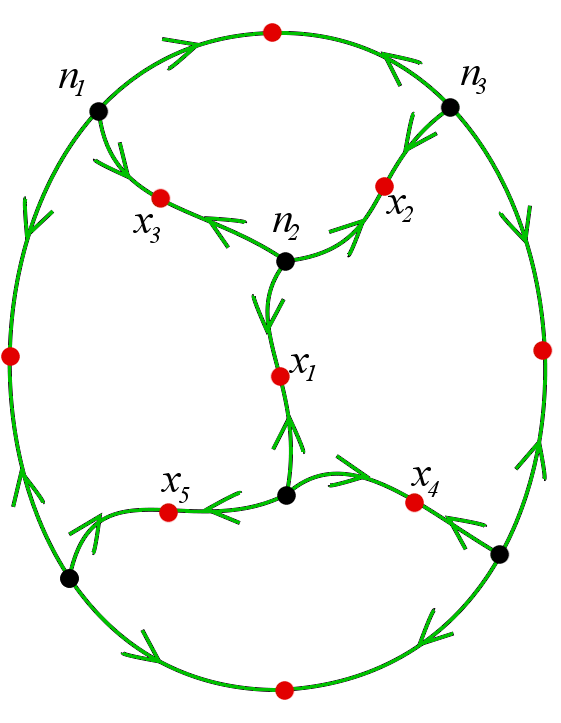}}
	\caption{With each graph we associate a squid graph by subdivision and reorientation of the edges.}
	\label{fig:squidgraph}
\end{figure}

\begin{df}
	Given a graph $\Gamma$ a \emph{squid graph} corresponding to it is $\gamma=(\s\Gamma,S)$, where $\s\Gamma$ is the graph \reef{GammaS} and $S$ is the set of squids $S= \{\lambda_n,\;n\in\T N\}$
\end{df}
Notice that a squid is a graph, however a squid graph is neither a squid nor a graph.

$\;$

Given several squid-graphs $\{\gamma_I\}_{I\in\T I}$ we will denote the disjoint sum of their squid-sets as
\be
	\T S=\bigsqcup_{I\in\T I}S_I
\ee

\subsection{From a squid-graph to 1-vertex foam\label{sc:1vfoam}}

Consider a squid graph $\gamma=(\s\Gamma,S)$. We want to construct a 2-complex, with precisely one internal vertex, whose boundary is the graph $\s\Gamma$ (which will be called a 1-vertex foam). We will do it by a formalization of the following shrinking procedure: draw the graph $\s\Gamma$ on a $3$-sphere of radius $1$ and then shrink the radius to zero. The track left by the graph defines the 2-complex $\kappa_\gamma$.

\begin{figure}[hbt!]
	{\centering
	\begin{tabular}{cc}
		{\begin{tabular}{c}
			\subfloat[A graph $\Gamma$]{\label{fig:constr_g}\includegraphics[width=0.25\textwidth]{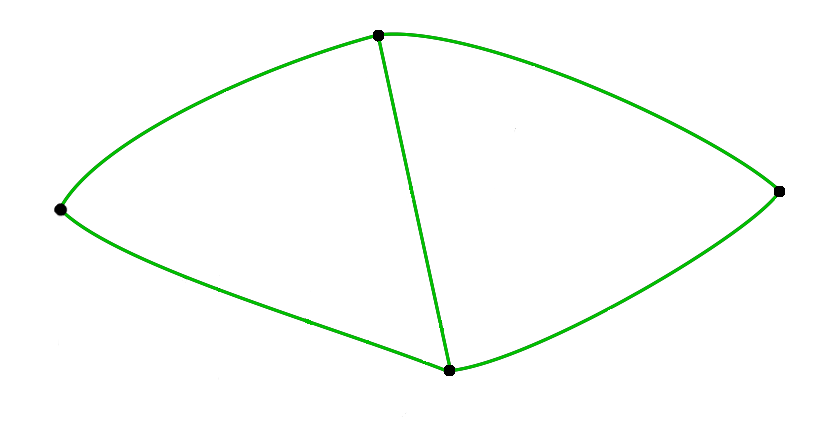} }\\
			\subfloat[The graph $\s\Gamma$ and the squid graph $\gamma$]{\label{fig:constr_sg}\includegraphics[width=0.25\textwidth]{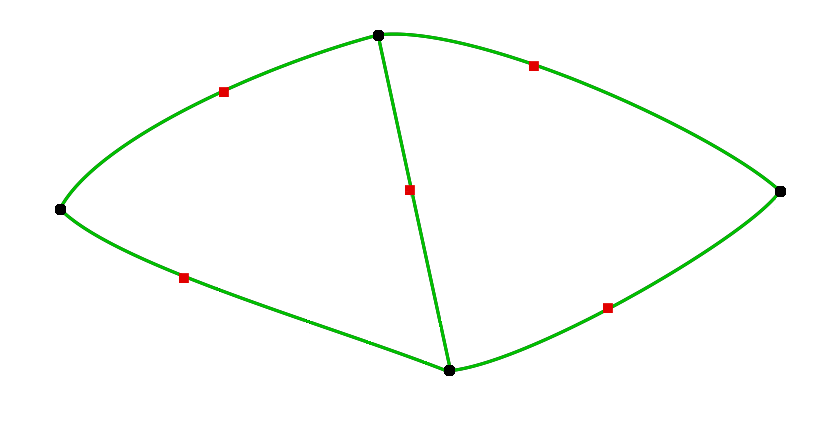} }
		\end{tabular}}
		&
		\begin{tabular}{c}
			$\;$\\
			\subfloat[A homotopy of the squid graph into a point $V$: the 1-vertex foam $\kappa_\gamma$]{\label{fig:constr_1vf}\includegraphics[width=0.50\textwidth]{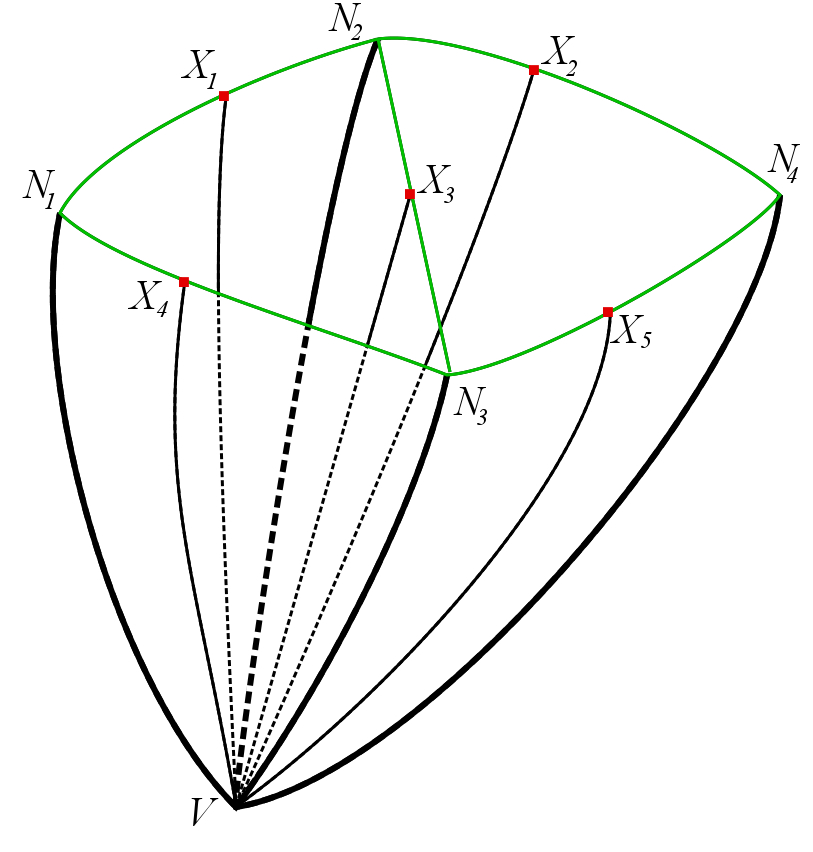} }
		\end{tabular}
	\end{tabular}
	}
	\caption{A graph (\ref{fig:constr_g}) is first turned into a squid graph (\ref{fig:constr_sg}) and next the spin foam is constructed as a homotopy of the squid graph to a point (\ref{fig:constr_1vf}). The thick edges are the traces of the nodes of the original graph $\Gamma$ (the head-nodes). The thin edges are the traces of the leg-nodes of the squid graph.}
	\label{fig:construction}
\end{figure}

A definition of the 2-cell complex $\kappa_\gamma$ follows quite clearly from \fref{construction}. Nonetheless we will spell out now a full rigorous definition consistent with the theory of 2-cell complexes \cite{Hatcher} by declaring its sets of faces, edges and vertices together with the gluing functions.

The graph $\s\Gamma$ viewed as a $\Delta$-complex is a triple $(\T L,\T N;f^{(0)}_{1\to0})$ where $f^{(0)}_{1\to0}:\pd\T L\to \T N$ is a function from boundaries of links to nodes (see the appendix \ref{sc:delta} for the details of notation). To form a 2-complex we need to add: one extra 0-simplex $v$ (the middle point of the sphere), a set of 1-simplexes which will be tracks of nodes $\T E_{\T N}:=\{I_n:n\in\T N\}$ and a set of 2-simplexes (faces), which will be tracks of links $\T F_{\T L}:=\{\Delta_\ell:\ell\in\T L\}$ (each $\Delta_\ell$ is a triangle).

The 2-complex we are constructing is given by a 5-ple $\kappa = \left(\T F,\T E,\T V; f_{2\to1},f_{1\to0} \right)$. The sets of 2-, 1- and 0-cells are, respectively, the following:
\be
	\T F = \T F_{\T L}\tab\T E=\T L\cup\T E_{\T N}\tab\T V = \{v\}\cup\T N
\ee
What one needs to define are the functions $f_{1\to0}:\pd\T E\to\T V$ and $f_{2\to1}:\pd\T F\to\bigsqcup\T E/\sim_{1}$ (where $\sim_1$ is the relation connecting preimages of function $f_{1\to0}$).

It is obvious that $f_{1\to0}$ restricted to $\T L$ is just $f^{(0)}_{1\to0}$, so lets take a look on the added edges. Each of $e\in\T E_{\T N}$ is labeled by a node $n\in\T N$ and all of them meet at the central point $v$. We orient them to be outgoing from $v$, so the function $f_{1\to0}$ acts like
\be
	f_{1\to0}:
	\left\{\begin{array}{rcl}
		\pd\T L\ni x &\mapsto&f^{(0)}_{1\to0}(x)\\
		\pd\T E_{\T N}\ni I_n(0) & \mapsto & v\\
		\pd\T E_{\T N}\ni I_n(1) & \mapsto & n
	\end{array}\right.
\ee
where $e:t\to e(t)$, ($t\in[0,1]$) is any parametrisation of an oriented 1-cell $e$.

The function $f_{2\to1}$ acts on a sum of boundaries of triangles. First lets take a look on a CW structure of a boundary of a single triangle $\Delta_{VNX}$. It is a 1-complex $\left(\{VN,NX,VX\},\{V,N,X\};f\right)$ (action of $f$ is obvious). The triangle $\Delta_{VNX}$ is a triangle with ordered vertices (we assume $V$ to be the first vertex, $N$ the second  and $X$ the third one), thus there is a natural orientation of its edges: from the earlier end to the later one (in the sense $s(VN)=V$, $s(NX)=N$, $s(VX)=V$). We introduce the $VNX$ structure at each of the triangles $\Delta_\ell$.

Given a link $\ell$ of the boundary graph $\s\Gamma$ we glue it to the $NX$-edge of the corresponding triangle $\Delta_\ell$ in a such way, that $N$ is always the starting point of the link and $X$ is its ending point. On the other hand edges $VN$ and $VX$ are glued to the internal edges i.e. the ones from the set $\T E_{\T N}$. The $V$ point will be the starting point of these edges, which, according to the $f_{1\to0}$ function, appear to be the $v$ vertex.

To be specific: function $f_{2\to1}$ act as:
\be\label{eq:f2to1}
	f_{2\to1}:\pd\Delta_\ell\ni\left\{
	\begin{array}{rcl}
		NX(t) &\mapsto&\left[\ell(t)\right]\\
		VN(t) & \mapsto & \left[I_{s(\ell)}(t)\right] \\
		VX(t) & \mapsto & \left[I_{t(\ell)}(t)\right] 
	\end{array}\right.
\ee
where square brackets stands for equivalence classes of $\sim_{1}$.

Few words of comment about the equivalence classes of $\sim_{1}$. Since the relation $\sim_{1}$ is the identification of ending points of edges, the equivalence class of $x\in{\mathrm{Int}}(\ell)$ is just $\{x\}$. When one considers the equivalence class of one of endpoints of edge $x\in\pd\T E$, it turns out to be the set $\left(f_{1\to0}\right)^{-1} \left(f_{1\to0}(x)\right)$ (for more details about $\sim_1$ see the appendix \ref{sc:delta}). To make formulas more transparent sometimes we will drop this relation. 

Notice that since all links $\ell$ starts at heads of the squids, each head is the $N$ vertex and each leg-node is a $X$ vertex. Thus two triangles can meet either by their $VN$ or $VX$ edges. Moreover since leg-nodes are always 2-valent, each $VX$ edge is shared by precisely 2 triangles, and since there are no 1-valent vertices, heads are at least 2 valent, so each $VN$ edge is shared by at least 2 triangles. The only boundary edges can be the $VX$ type ones. In fact all of them are the boundary of this foam.

Since the boundary of $\kappa_\gamma$ is just $\s\Gamma$, it can be decomposed into the same squids, as the original graph $\Gamma$. We will keep it in mind by adding the set $S$ to the object we have just obtained. So the final 1-vertex foam will be denoted as
\be
	(\kappa_\gamma,S)
\ee
Collection of such 1-vertex foams (one per each $\gamma\in\K G$) is \emph{the base} of our inductive construction.

\subsection{Gluing along the squids\label{sc:gluing}}

Consider two 1-vertex foams $\kappa_{\gamma_1}$ and $\kappa_{\gamma_2}$ coming from squid graphs $\gamma_1$ and $\gamma_2$ together with their sets of squids $S_1$ and $S_2$. Suppose that a squid $\lambda\in S_1$ on the boundary of $\kappa_{\gamma_1}$ is homeomorphic to a squid $\lambda'\in S_2$ on a boundary of $\kappa_{\gamma_2}$ (i.e. they have the same number of legs). We will define now the gluing of $\kappa_{\gamma_1}$ and $\kappa_{\gamma_2}$ along the squids $\lambda$ and $\lambda'$. The operation will be denoted by $\kappa=\kappa_{\gamma_1}\cup_{(\lambda,\lambda')}\kappa_{\gamma_2}$. To be more specific we denote the duality map by $\phi:\lambda\to\lambda'$ defining the morphism of these squids (it needs to be a bijection).

\begin{figure}[hbt!]
	\centering
	\begin{tabular}{cc}
		\begin{tabular}{c}
			\subfloat[The boundary squid graphs $\gamma_1$ and $\gamma_2$ of two 1-vertex  spin foams $\kappa_{\gamma_1}$ and $\kappa_{\gamma_2}$. The dashed lines mean that only a part of each graph is depicted, $\lambda$ and $\lambda'$ denote squids.]
					{\label{fig:gluing_1}\includegraphics[width=0.45\textwidth]{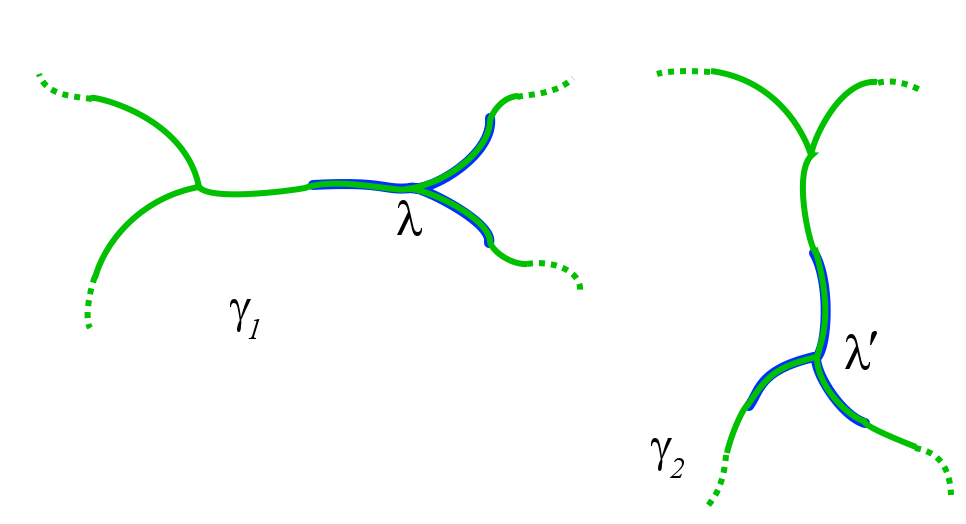} }\\
			\subfloat[The foam $\kappa_{\gamma_1}$ bounded by the squid graph $\gamma_1$]
					{\label{fig:gluing_2}\includegraphics[width=0.45\textwidth]{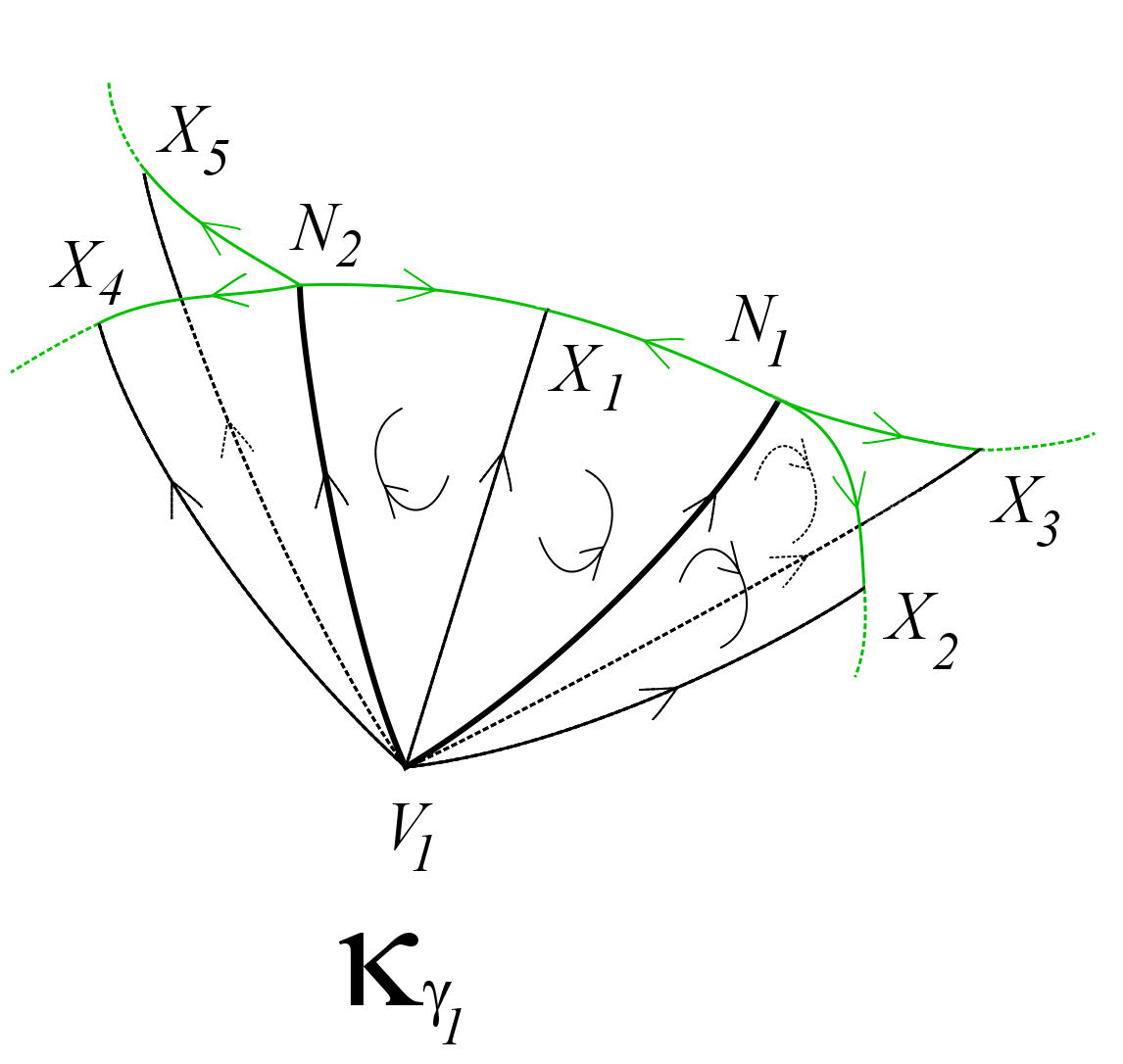} }
		\end{tabular}
		&
		\begin{tabular}{c}
			$\;$\\
			\subfloat[Glueing of the two foams along the squids $\lambda$ and $\lambda'$]{\label{fig:gluing_3}\includegraphics[width=0.45\textwidth]{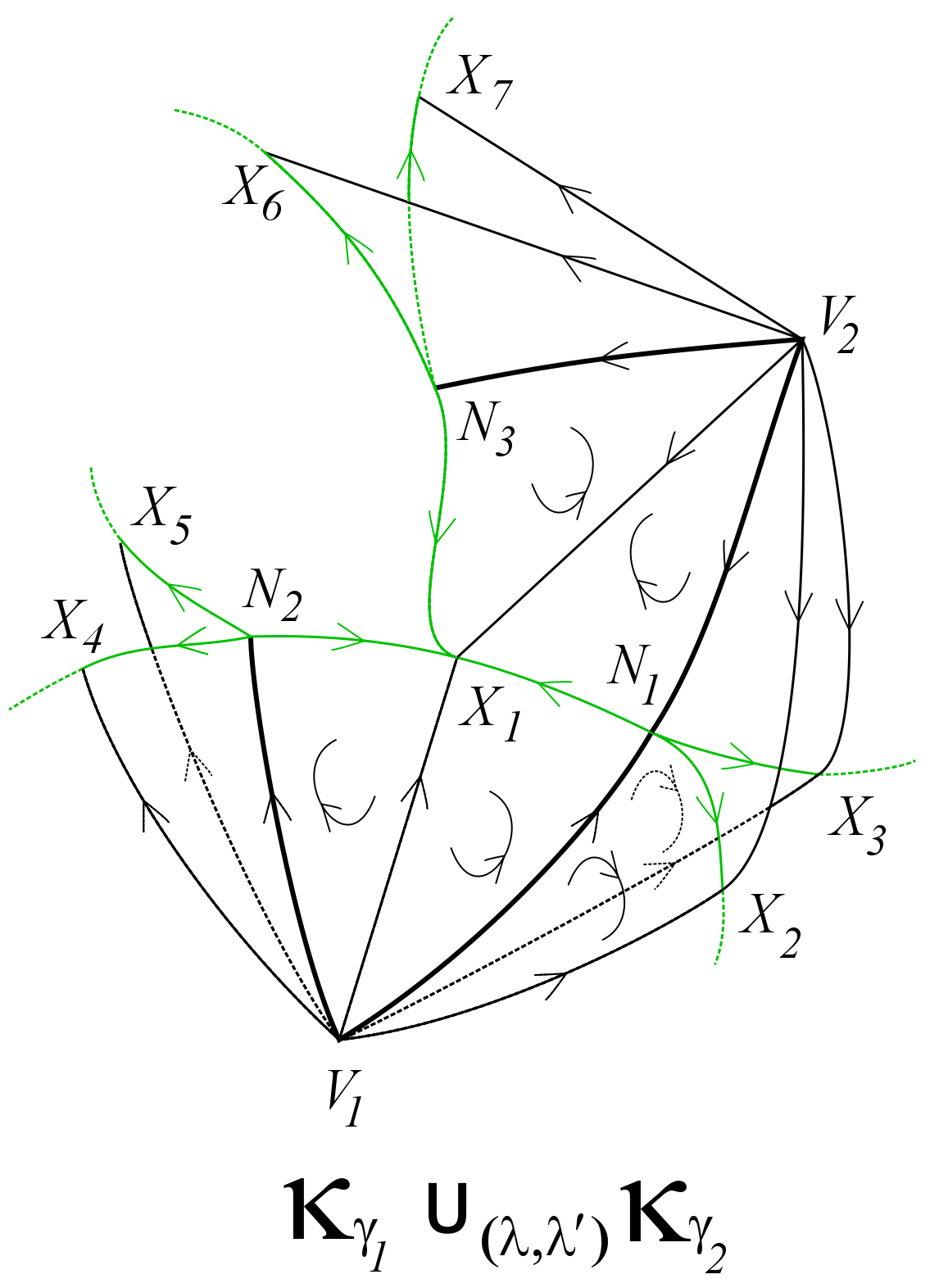} }
		\end{tabular}
	\end{tabular}
	\caption{The two initial graphs $\Gamma_1$ and $\Gamma_2$ are glued along the squids $\lambda$ and $\lambda'$ (\ref{fig:gluing_1}) into the complex $\kappa_1\cup_{(\lambda,\lambda')}\kappa_2$.}
	\label{fig:gluing}
\end{figure}

The 2-complex $\kappa=\kappa_1\cup_{(\lambda,\lambda')}\kappa_2$ is just the union of the complexes $\kappa_1$ and $\kappa_2$ with certain pairs of edges identified\footnote{We are referring to the links of the squids as to the edges of the foams.}. The figure \ref{fig:gluing} shows an example of the gluing. To be more specific:
	\begin{itemize}
		\item Lets take a disjoint union of the complexes: $\tilde\kappa=\kappa_{\gamma_1}\sqcup\kappa_{\gamma_2}$.
		\item The map $\phi$ between $\lambda$ and $\lambda'$ gives a set of pairs of edges:
			\be
				R_\phi=\left\{\alpha = (\ell,\ell')\;:\; \ell\in\T L_{\lambda}\;,\; \ell'\in\T L_{\lambda'}\;,\;\ell'=\phi(\ell)\right\}
			\ee
		\item We identify each pair of edges $\alpha\in R_\phi$:
			\be
				\tilde\kappa\mapsto\tilde\kappa/\alpha_1\mapsto\left(\tilde\kappa/\alpha_1\right)/\alpha_2\mapsto
					\cdots\mapsto\kappa
			\ee
			
			Identifying a pair of edges is a procedure after which the resulting 2-complex differs from the original one only by the fact, that two edges became one edge (and by all topological implications of it). The detailed definition of this operation is given in the appendix \ref{sc:cwgluing} together with the theorem, that the result of such gluing does not depend on the order of gluings (app. \ref{sc:theorem}). Thanks to that theorem our operation is well defined.
	\end{itemize}

	The resulting 2-complex will be denoted as	
		\be
			\kappa=\kappa_{\gamma_1}\cup_{(\lambda,\lambda')}\kappa_{\gamma_2}=\left(\T F, \T E, \T V; f_{2\to1}, f_{1\to0}\right)
		\ee
	and its components are as follows:
	\begin{itemize}
		\item The set of faces is just the union: $\T F=\T F_1\cup\T F_2$.
		
		\item The set of edges is the union divided by a relation: $\T E=\T E_1\cup\T E_2/_{\sim_{\phi,1}}$, where two edges $e$, $e'$ are in the relation $\sim_{b,1}$ if and only if $e\in\lambda$, $e'\in\lambda'$ (or opposite) and $\phi(e)=e'$ (or $\phi^{-1}(e)=e'$ in the opposite case).
		
		\item The set of vertices is defined in analogous way: $\T V=\T V_1\cup\T V_2/_{\sim_{\phi,0}}$ where $v\sim_{\phi,0} v'$ if and only if $v\in\lambda_1$ and $v'\in\lambda_2$ (or opposite) and $\phi(v)=v'$ (or $\phi^{-1}(v)=v'$ in the opposite case).
		
		\item The function $f_{1\to0}$ coincides with the functions $f_{1\to0}^{(1)}$ and $f_{1\to0}^{(2)}$ at their domains, followed by the projection $\pi_{\sim_0}$ onto the equivalence classes of the relation $\sim_{\phi,0}$:
			\be
				f_{1\to0}:\left\{\begin{array}{rcl}
						\pd\T E_1\ni x &\mapsto& \pi_{\sim_0}\circ f^{(1)}_{1\to0}(x)\\
						\pd\T E_2\ni x &\mapsto& \pi_{\sim_0}\circ f^{(2)}_{1\to0}(x)
					\end{array}\right.
			\ee
			however one needs to do the consistency check with the relation $\sim_{\phi,1}$, i.e. check if $x\sim_{\phi,1} x'$ implies $f_{1\to0}(x)=f_{1\to0}(x')$?
		 	
			Outside the glued squids it is obviously satisfied, since in this regime equivalence classes of $\sim_{\phi,1}$ are one-element sets. Assume thus, that $x\in\pd e$ for $e\in\lambda$ and we have $x'\not=x$ such that $x'\sim_{\phi,1}x$. If it is so, $x'$ must be in $\lambda'$, and $\phi(x)=x'$. We have $f_{1\to0}(x')=\pi_{\sim_0}\circ f^{(1)}_{1\to0}(x)$ and $f_{1\to0}(x)=\pi_{\sim_0}\circ f^{(2)}_{1\to0}(x')$. However since $\phi$ is a morphism of $\Delta$-complexes, the condition $\phi(x)=x'$ must follow $\phi(f^{(1)}_{1\to0}(x))=f^{(2)}_{1\to0}(x')$, thus $f_{1\to0}(x)=f_{1\to0}(x')$, what ends the proof.
		
		\item The function $f_{2\to1}$ coincides with the functions $f_{2\to1}^{(1)}$ and $f_{2\to1}^{(2)}$ at their domains, followed by the projection $\pi_{\sim_1}$
			\be
				f_{2\to1}:\left\{\begin{array}{rcl}
						\pd\T F_1\ni x &\mapsto&\pi_{\sim_1}\circ f^{(1)}_{2\to1}(x)\\
						\pd\T F_2\ni x &\mapsto&\pi_{\sim_1}\circ f^{(2)}_{2\to1}(x)
					\end{array}\right.
			\ee		
	\end{itemize}

The new set of boundary squids of $\kappa_{\gamma_1}\cup_{(\lambda,\lambda')}\kappa_{\gamma_2}$ is
\be
	S=\left(S_1\cup S_2\right)\setminus\{\lambda,\lambda'\}
\ee
Thus we have just obtained a foam with squid structure on its boundary $(\kappa,S)$ being the gluing of two 1-vertex foams $(\kappa_{\gamma_1},S_1)$ and $(\kappa_{\gamma_2},S_2)$.

\subsection{Continuation of gluing procedure to more general cases}

What need to be done now is to show, that the same step we have just done from $n=1$ to $n+1=2$ can be done from arbitrary $n$ to $n+1$ in inductive way.

The key step of the construction is noticing that all we needed during our construction so far was the knowledge, that the foam we glue is the proper spin foam with squids drawn on its boundary. 
Indeed: none of the steps in the subsection \ref{sc:gluing} requires, that the complex $\tilde\kappa$, whose cells were glued, was of the form $\kappa_{\gamma_1}\sqcup\kappa_{\gamma_2}$. The only thing that is needed are general properties of $2-\Delta$-complex and the squid structure we introduced on the boundary i.e. decomposition into squids. Thus any result of gluing procedure may be a starting point of of another gluing procedure of this type. Moreover the gluing procedure is independent on the order in which pairs of squids are glued, what is obvious implication of the commutativity of gluing of pairs of edges (app. \ref{sc:theorem}).

So finally are able to complete the construction.

\begin{enumerate}
	\item Let $(\T G, \T R)$ be a graph diagram. Let $\K G$ be a set of squid graphs constructed from graphs being elements of $\T G$. Let $\K R$ be the relation on the set of all squids defined as follows: $\lambda$ is in relation with $\lambda'$ iff the head of $\lambda$ is in  relation $\T R_{\rm node}$ with the head of $\lambda'$. For each pair $(\lambda,\lambda')\in\K R$ the link relation $\T R_{\rm link}$ induces naturally a morphism of 1-complexes $\phi:\lambda\to\lambda'$ which identifies each leg of $\lambda$ with a leg of $\lambda'$ in 1-1 way.
	
	\item We construct a family of 1-vertex foams $\kappa_\gamma$ creating one from each squid graph $\gamma\in\K G$ (i.e. from each $\Gamma\in\T G$)
	
	\item We take the disjoint union of all these 1-vertex foams:
		\be\label{eq:tildekappa}
			\tilde\kappa=\bigsqcup_{\gamma\in\K G}\kappa_{\gamma}
		\ee
	
	\item We denote $\kappa_0=\tilde\kappa$, then we order pairs of squids $r\in \K R$ by numbers from $n=1$ to $\#\K R$ and perform gluings by saying $\kappa_n=\kappa_{n-1}/\sim_{r_n}$, i.e. we glue the complex along the pairs of squids one after the other.
	\item The resulting 2-complex is
		\be\label{eq:kappa.final}
			\kappa=\kappa_{\#\K R}
		\ee
\end{enumerate}


\section{Properties of such foam\label{sc:properties}}

In the previous section have constructed a 2-complex $\kappa$ corresponding to a squid graph diagram $\K D= (\K G,\K R)$. Now we will analyse the structure of this foam. We will discuss the way faces may intersect at edges, edges may intersect at vertices and we will give some examples of possible topologies of the faces.

\subsection{Types of edges\label{sc:TypesOfEdges}}

Using the notation introduced in \sref{1vfoam} (see \reef{f2to1}) one may distinguish three types of edges: $VN$, $VX$, and $NX$, however the $NX$ edges may be of two subtypes: internal or external. All these types are presented at the figure \ref{fig:properties}. Properties of the edges are:
\begin{figure}[hbt!]
	\centering
	\includegraphics[width=0.6\textwidth]{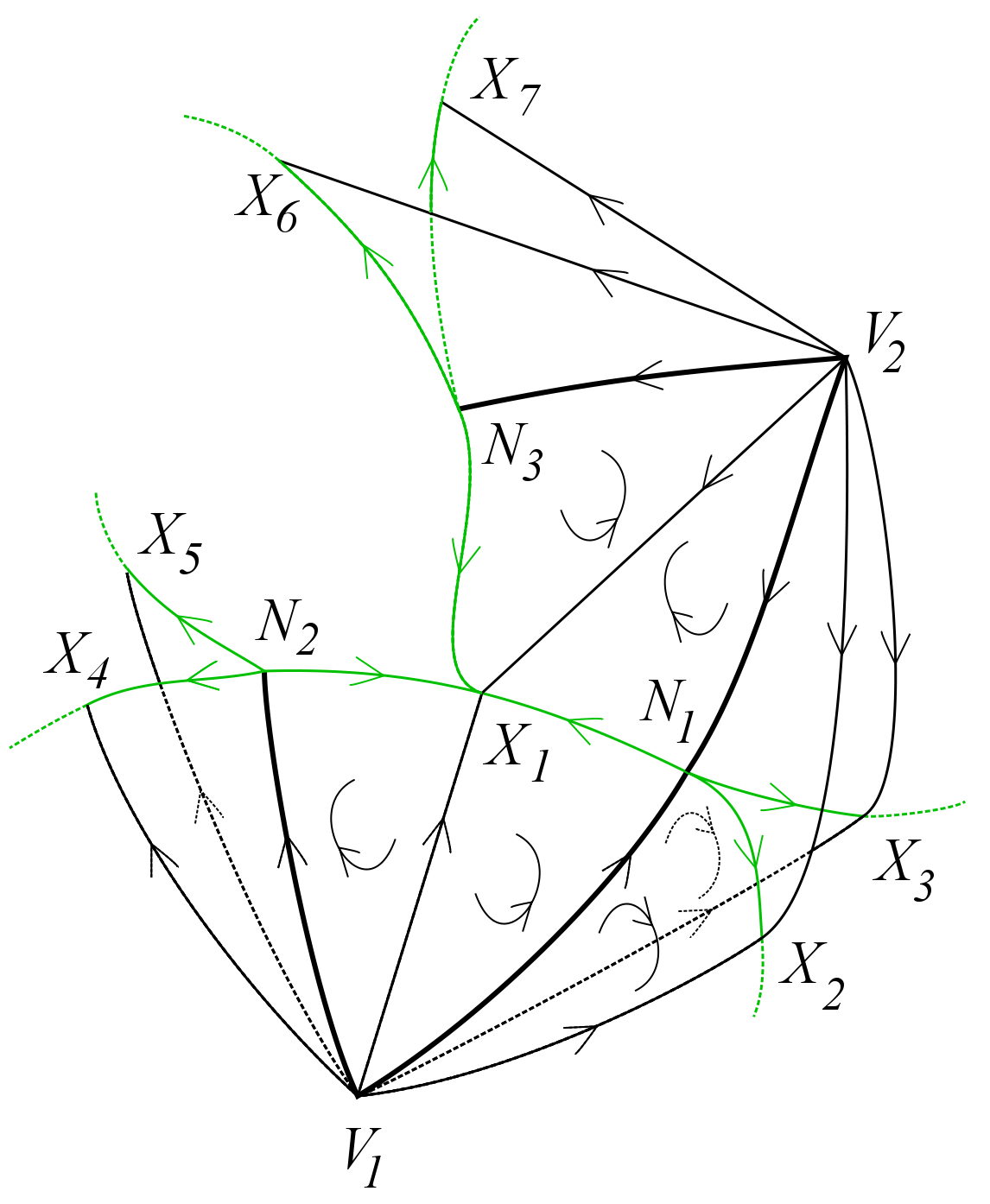}
	\caption{A fragment of a spin foam. The thick black edges are of the type $VN$, the thin black ones are the type $VX$, and the green ones are $NX$-type edges. The orientations of the faces are always consistent with those of the $NX$-edges. Some of $NX$-edges are boundary edges.}
	\label{fig:properties}
\end{figure}

\begin{enumerate}
	\item The $VN$-type edge is a history of the \emph{head} of a squid (the head itself correspond to the point $N$). It is always sheared by at least two faces (actually: to as many faces, as many legs had the squid it was build from). The faces are always consistently oriented with the edge.
		
	\item The $VX$-type edge is a history of the \emph{leg-node} of a squid (the leg node itself correspond to the point $X$). It is always sheared by precisely two faces (coming from the links that were meeting at the node). Both these faces are oriented opposite to the edge.

	\item The $NX$-type edge is a leg of a squid. It is a boundary edge if and only if the squid it belongs to was not glued to another squid.
	
	\item The $NX$-type edge is an internal edge of a complex if and only if the squid it belongs to was glued to another squid. In such case this edge is sheared by precisely two faces. Orientation of the faces is consistent to the orientation of the edge (by definition, see \ref{sc:1vfoam}).
	
\end{enumerate}

The edges of type $VX$ and the internal edges of type $NX$ (i.e. type 2 and type 4) will be called \emph{removable} and actually will be removed in the \sref{Removing}.

\subsection{Types of vertices\label{sc:TypesOfVertices}}

There are three main types of vertices in 2-complex of our construction: $V$, $N$ and $X$. The $N$ and $X$ type split however into respectively two and three subtypes, so we have six classes of vertices to describe.

\begin{enumerate}
	\item Vertices of type $V$. They are always internal vertices. There is one such vertex for each squid graph $\gamma\in\K G$.
	
	\item Vertices of type $N$ are the heads of squids. Such vertex is an internal vertex if and only if the squid it came from was glued with another squid. In such case the $N$-type vertex looks like the the vertex of type $V$ coming from the $\theta$-graph (see \fref{BandC} - of course the similarity is only local).
		
		There are always two edges of type $VN$ ending at such a vertex and a number of $NX$ type edges starting at this vertex. Since $NX$ edges are removable, after removing them the $N$ type vertex becomes a bivalent vertex in the middle of an $VV$ edge, thus it is also called \emph{removable}.
	
	\item Vertex of type $N$ coming from the non-glued squid is the boundary vertex. It is then a node of the boundary graph.
	
	\item Vertices of type $X$ are the leg-nodes of the squids. Such vertex is an internal vertex if an only if the squid leg it came from is a half of a link that belongs to a cyclic equivalence class of the face relation $\T R_{\rm face}$ or a 1-element equivalence class of $\T R_{\rm link}$ (i.e. it is an element of an equivalence class of $\T R_{\rm face}$ relation of type 2, 3(b) or 4(b) see \sref{FaceRelation})
Locally it looks like an $V$-type vertex for a loop graph (see fig. \fref{BandC})
		
		Since all edges ending at such $X$-type vertex are removable (i.e. $VX$ and internal $NX$), the vertex itself will be also called \emph{removable}.
	
	\begin{figure}[hbt!]
		\centering
			\subfloat[A fragment of a spin foam containing  internal vertices $N_2$ and $X$.]{\label{fig:BandC_foam}\includegraphics[width=0.4\textwidth]{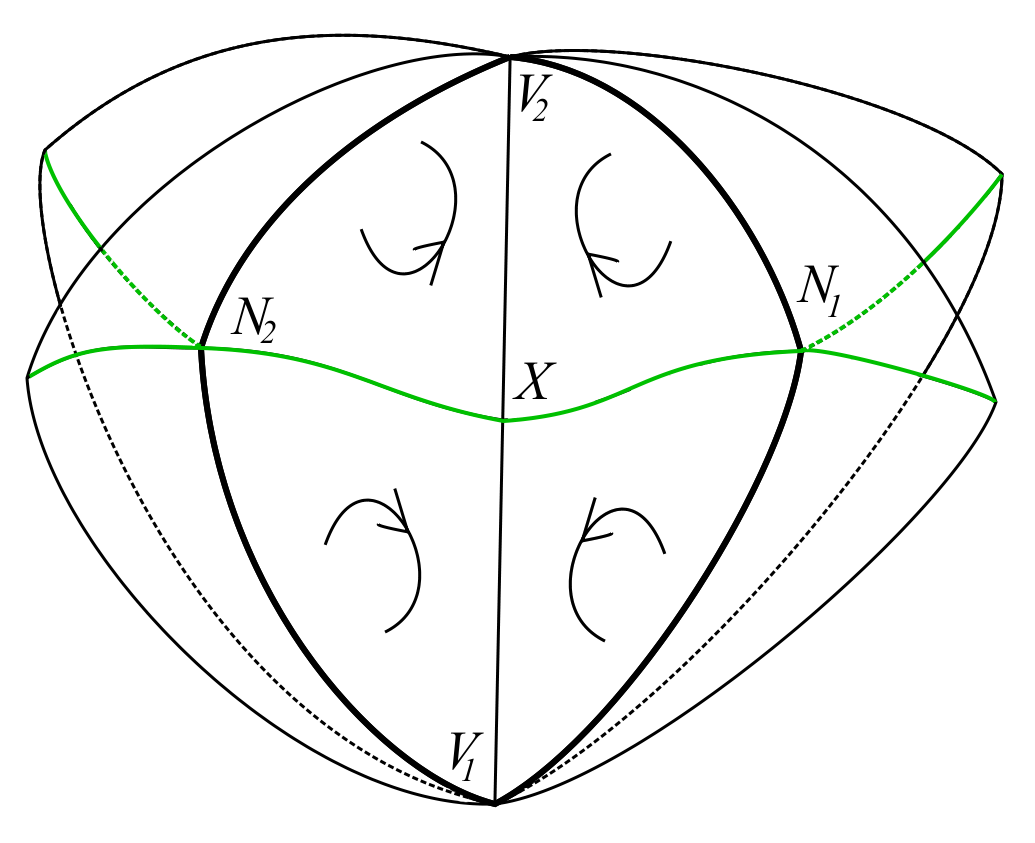} }
			\subfloat[Neighborhoods of the vertices  $N_2$ and, respectively, $X$. The neighborhoods are bounded by the depicted  graphs.]{\label{fig:BandC_networks}\includegraphics[width=0.4\textwidth]{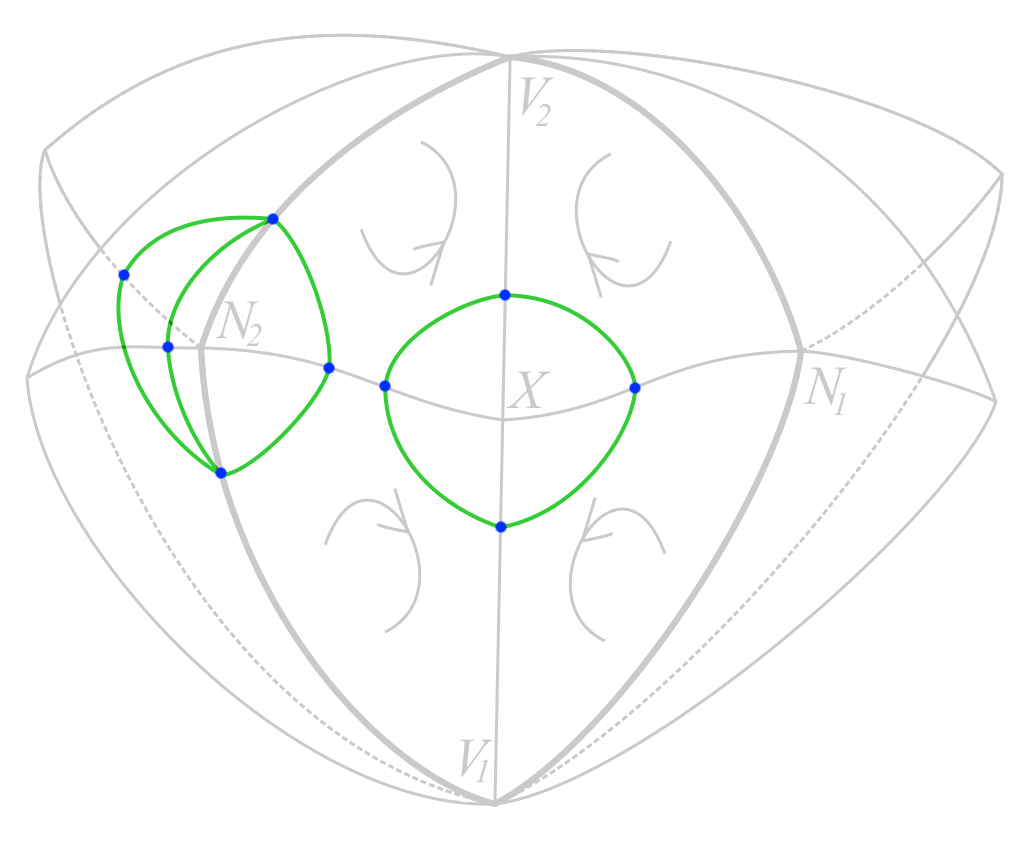} }
		\caption{The spin foam (a) arises from the foam of fig.\ref{fig:properties} by gluing the squids of the heads $N_2$ and, respectively, $N_3$. We focus on the foam vertices $N_2$ and  $X$. Their neighborhoods  (b) are bounded by a $\theta$-like
		graph  ($N_2$) and, respectively, a loop-like graph ($X$). The neighborhood of $X$ is a disc.}
		\label{fig:BandC}
	\end{figure}

	\item If none of the squids that a vertex of type $X$ belongs to is glued to any other squid, it is a simple boundary vertex
	(it is a leg node in the middle of a link being unrelated to any other link by $\T R_{\rm link}$ relations, i.e. belonging to a type 1 equivalence class of the face relation $\T R_{\rm face}$).
		
		Removing of the $VX$ edge ending at such a vertex makes it a boundary bivalent node. It will be called \emph{removable}.

	\item The last possibility for $X$-type vertex is that it is the middle of a link $\ell$ being an element of an open equivalence class of the face relation $\T R_{\rm face}$ (i.e. an equivalence class of type 3(a) or 4(a) - see \sref{FaceRelation}).
		
		Such a vertex is also \emph{removable}, because the internal edges ending at it are $VX$ or internal $NX$ type, and after removing them the vertex becomes a bivalent boundary node, like in the previous case.

\end{enumerate}

Thus the only non-removable vertices are type $V$ and boundary type $N$ vertices.

It is worth to notice that a neighbourhood of each vertex of type $X$ is topologically a disc. There are two possibilities: for a boundary vertex $X$, and for an internal vertex $X$ (see \fref{Cvertices}). In both cases the edges meeting at $X$ form a sequence of $NX$- and $VX$-type edges alternately i.e. a $NX$ edge is followed by an $VX$ edge and an $VX$ edge is followed by a $NX$ edge. In case of a boundary vertex $X$ the sequence starts and ends with two (different)  boundary $NX$-edges. In case of an internal vertex $X$ we can choose any $NX$ edge as a starting one and the last $VX$ edge in a sequence is followed by the beginning $NX$ edge.

\begin{figure}[hbt!]
		\centering
			\subfloat[A boundary  vertex $X_1$. The depicted segment of the boundary is $N_1X_1N_2$.]{\includegraphics[width=0.4\textwidth]{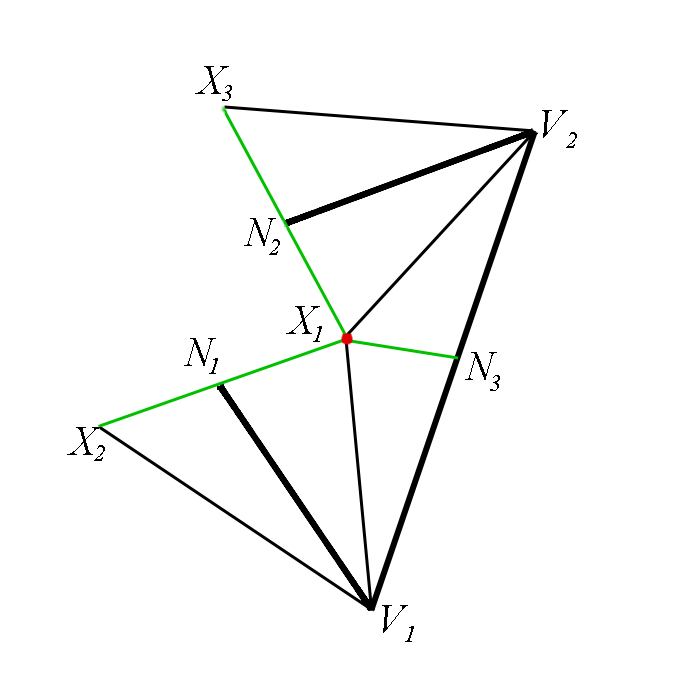} }
			\subfloat[An internal vertex  $X$]{\includegraphics[width=0.4\textwidth]{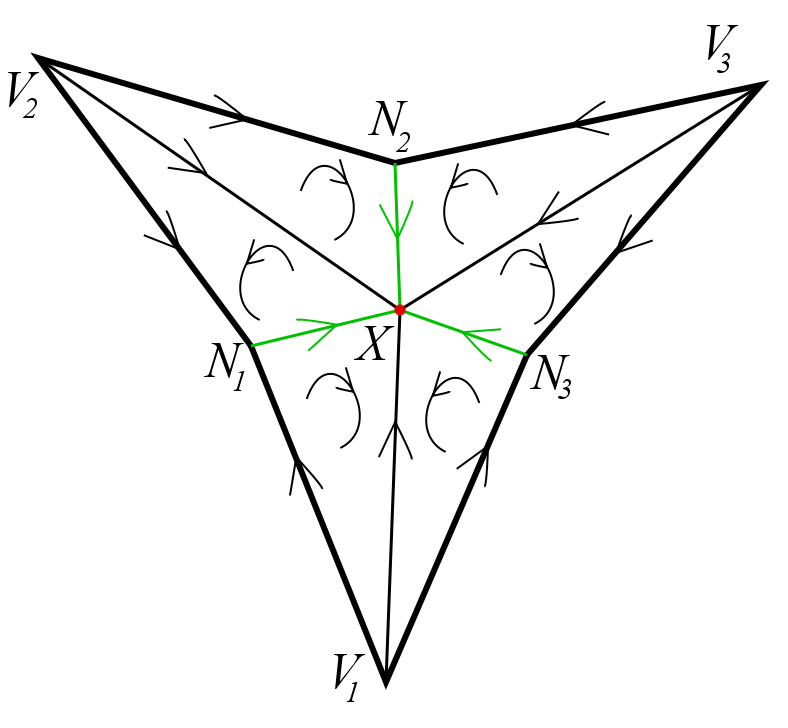} }
		\caption{Boundary and internal  $X$-vertices}
		\label{fig:Cvertices}
\end{figure}

\subsection{Description of faces}

The set of faces does not change during the gluing procedure, so final set $\T F$ is just the union of original sets $\T F_{\T L_\gamma}$ for all initial squid graphs $\gamma\in\K G$. The faces are all triangular. Each face has two internal edges ($VN$ and $VX$), and the third edge ($NX$ type) is also internal, if the squid the face came from was glued to another squid, and it is boundary edge, if the squid was not glued to anything. Topologically each face is a disk, placed onto some skeleton. Given a face, none of its edges is glued with other edges of the same face (see \sref{TypesOfEdges}).


\section{The final operator spin foam corresponding to an operator spin network diagram\label{sc:coloring}}

\subsection{Removing of redundant edges and vertices\label{sc:Removing}}

In previous section some edges and vertices have been marked as removable. These were the $VX$ edges, the internal $NX$ edges, $X$ vertices and internal $N$ vertices. They were auxiliary in our construction, while the other edges and vertices have a direct correspondence with elements of a graph diagram. We will remove them now from the 2-complex $\kappa$ of the equation \reef{kappa.final} by merging the higher dimension cells sharing the removable ones.

The resulting 2-complex $\kappa_{\T D}$ can be characterised in terms of the corresponding graph diagram $\T D=(\T G,\T R)$:
\begin{itemize}
	\item for each graph $\gamma\in\T G$ there is one internal vertex $v_\gamma$. 
	\item for each \emph{boundary} node $n$ of the graph diagram (i.e. a node that is unrelated by the node relation) there is a boundary vertex of the 2-complex (denoted also by $n$).
	\item for each equivalence class of the edge relation $\T R_{\rm edge}$ there is an internal edge of the 2-complex. If the equivalence class is one-element $\{n\}$, then the edge meets the boundary (at the boundary vertex corresponding to the node $n$), and ends at the internal vertex corresponding to the graph that $n$ belongs to. If the equivalence class is two-element $\{n,n'\}$, then the edge connects the internal vertices corresponding to the graphs that $n$ and $n'$ belong to.
	\item for each link of the \emph{boundary} graph of the graph diagram there is a boundary edge of the 2-complex. The edge connects the boundary vertices of the 2-complex that correspond to the same nodes of the boundary graph, that the link connects.
	\item for each equivalence class of the face relation $\T R_{\rm face}$ there is an oriented face of the 2-complex. It is oriented and glued to the skeleton of the 2-complex in a way that will be described shortly (see \sref{MergedProperties}).
\end{itemize}

\subsection{Properties of the final 2-complex\label{sc:MergedProperties}}
In the \sref{properties} we have discussed the properties of vertices, edges and faces of the 2-complex obtained out of squid graphs, before removing the extra cells. Now we will characterise possible classes of cells resulting after the removing.

\subsubsection{Vertices}
Each internal vertex of the resulting 2-complex comes from a vertex of type $V$. Its structure is completely characterised by the graph $\Gamma\in\T G$ it corresponds to (the graph is the boundary of a sufficiently small neighbourhood of the vertex).

Each boundary vertex of the resulting 2-complex comes from a node of the graph diagram unrelated to any other nodes by $\T R_{\rm node}$ (i.e. from a boundary vertex of type $N$). Its structure is given by the structure of the node of the boundary graph it corresponds to (a sufficiently small neighbourhood of a boundary vertex $N_n$ is a Cartesian product of a squid $\lambda_n$ and an interval $[0,1[$).

\subsubsection{Edges}
There are three types of edges: the boundary edges, the internal edges with one end on the boundary and the internal edges with no end at the boundary (there are no internal edges with both ends at the boundary).

Each boundary edge comes from merging of two boundary $NX$ edges shearing the $X$ vertex.

The internal edges with one end at the boundary correspond to one-element equivalence classes of the edge relation $\T R_{\rm edge}$. Each of them comes from a single $VN$ edge, where $N$ is on a boundary.

Each internal edge with no end at the boundary comes from a pair of $VN$ edges shearing the $N$ vertex (thus the $N$ vertex was removed). It is possible, that both merged $VN$ edges started at the same $V$ vertex. In such a case obtained internal edge is a loop starting and ending at one internal vertex. However in general the internal edges connect pairs of internal vertices.

\subsubsection{Faces}

Each face of the final 2-complex corresponds to one equivalence class of the face relation $\T R_{\rm face}$. Thanks to the structure of the $X$ vertices each face is a union of the triangular faces of the complex $\kappa$ \reef{kappa.final} shearing one $X$-type vertex. Therefore types of faces correspond to the types of equivalence classes of $\T R_{\rm face}$ and to the types of (removed) $X$-vertices (see \sref{FaceRelation} and respectively \sref{TypesOfVertices}).

There are two types of faces: faces which overlap the boundary edges and faces which overlap only the internal edges.

\begin{itemize}
	\item Each face which overlaps the boundary edges corresponds to an open equivalence class of $\T R_{\rm face}$ (i.e. the equivalence class of type 1, 3a) or 4a)). The $X$-vertex it is coming from was a boundary vertex (i.e. $X$ vertex of type 5 or 6), thus the face contains precisely one boundary edge (being the boundary $N_1N_2$ link that came from the same $X$ vertex). We orient this face in agreement with the boundary link it contains. Other edges of that face are, in order: the $X_2V_1$ (where $X_2$ is the ending of the boundary edge), then possibly some sequence of edges $V_1V_2,\ldots,V_{k-1}V_k$ (however, $k$ may be equal to 1), and then $V_kN_1$.
	
	Some of the $V_i$s may be equal, in such a case it effects the topology of the face. Moreover it may happen that $N_1=N_2$ (the boundary edge is a loop), and thus $V_1=V_k$, in such a case all the edges $N_1V_1$ and $N_2V_k$ are equal - with all the consequences for the topology (i.e. the face is either a cone or a cylinder).
	
	It is impossible to obtain a face that contains more then one boundary edge.

	\item Each face which overlaps no boundary edges corresponds to a closed equivalence class of $\T R_{\rm face}$ (i.e. the equivalence class of type 2, 3b) or 4b)). The $X$-vertex it is coming from was an internal vertex (i.e. $X$ vertex of type 4). All edges of this face are internal $VV$ edges.
	
	To orient such a face recall the structure of the $X$ type vertex. In previous section we have not used the orientation of the links of unsquided graphs, but we will invoke it now (as in the previous item). Each $VNX$ triangle meeting at considered $X$ vertex inherits an orientation from the unsquided graph. One can check that for each two triangles neighbouring at this $X$ vertex their orientations agree. Therefore the face obtained by removing the $VX$ and $NX$ edges also inherits that orientation.
	
	In other words we orient the faces in such a way that if one considers a neighbourhood of any internal vertex $v_\Gamma$, then its boundary agrees with the graph $\Gamma$, including the orientation.
	
	The edges of such face form a sequence. Some elements of this sequence (edges or vertices) may appear more then once.
\end{itemize}

Notice that however the interior of each face is a disc, its boundary may be glued in a topologically nontrivial way. An example of it is shown and explained at \fref{p-plane}.

\begin{figure}[hbt!]
	\centering
	\begin{tabular}{cc}
		\subfloat[The fragments of squid graphs to be glued. Dashed green lines express that only fragment of graphs are shown. Dashed black lines show the node relation, dotted black lines together with small letter $a,a',\ldots,d,d'$ show the link relation.]{\label{fig:p-plane_1}\includegraphics[width=0.4\textwidth]{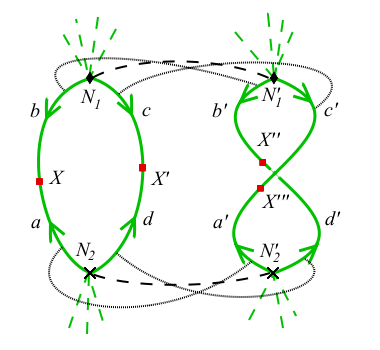} }&
		\subfloat[The segments of the 1-vertex foams bounded by the squides. For simplification they  have been cut along the edges $V_1X'$ and $V_2X''$]{\label{fig:p-plane_2}\includegraphics[width=0.4\textwidth]{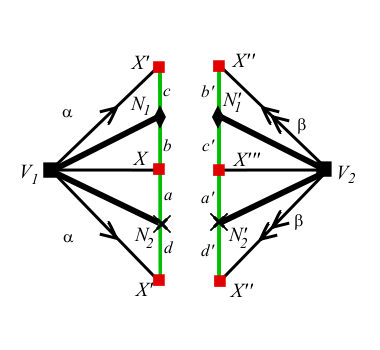} }
		\\
		\subfloat[The previous picture without the simplification. The orientations of the green edges show, the way they will be glued with the primed ones]{\label{fig:p-plane_3}\includegraphics[width=0.4\textwidth]{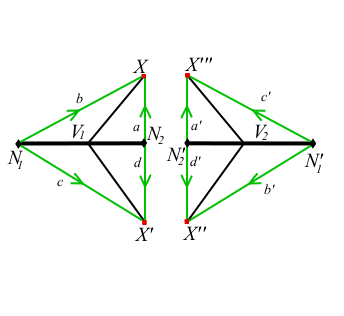} }&
		\subfloat[The result of the glueing. The arrows define the way  the points on the sides of the square are identified. The only  edges of the 2-complex are the two $V_1V_2$ edges forming the ''equator'' of the projective plane (The points $N_1$, $N_2$ and $X$ are not any more vertices of the 2-complex).]{\label{fig:p-plane_4}\includegraphics[width=0.4\textwidth]{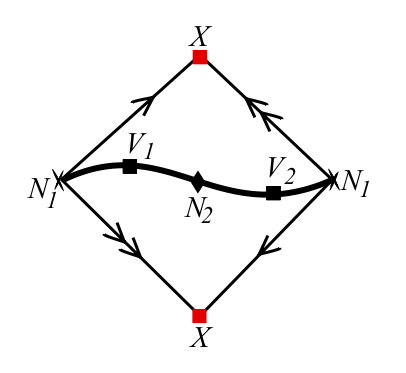} }
	\end{tabular}
	\caption{An example of  a face having the projective plane topology. At each step primes shows objects that will be identified in later steps.}
	\label{fig:p-plane}
\end{figure}

\subsection{The coloring}
Having defined the 2-complex $\kappa_{\T D}$ in \sref{Removing} for the graph diagram $\T D=(\T G,\T R)$ now we will define the operator spin foam $(\kappa_{\T D},\rho,P,A)$ for the operator spin network diagram $(\T G,\T R;\rho,P,A)$. To define it we need to define the coloring of $\kappa_{\T D}$, which will be induced by the coloring of the diagram in a straightforward way:

\begin{itemize}
	\item Each face $f$ corresponding to the equivalence class $[\ell_i]=\{\ell_1,\ldots\ell_k\}$ of the relation $\T R_{\rm face}$ is colored by the representation $\rho_f:=\rho_{\ell_i}$ for an on (arbitrary) representative of the equivalence class (because the coloring $\rho_\ell$ is constant on the equivalence classes). The corresponding carrier Hilbert spaces will be denoted by $\Hil_{\rho_f}$.
		
		This coloring induces the coloring of the boundary edges in a way consistent with the coloring of the boundary graph of the operator spin network diagram.

	\item Each edge $e$ which has one end on the boundary corresponds to a boundary node $n$ of the diagram. Each such node is colored by an operator $P_n$ (see \reef{nodeoperator}), which induces a coloring of the edge $P_e:=P_n$.
		
	\item Each edge $e$ which has no end on the boundary corresponds to a pair of related nodes $\{n,n'\}$ in the diagram. Each such pair is colored by $P_{\{n,n'\}}$ (see \reef{P.nn}), which induces a coloring of the edge $P_e:=P_{\{n,n'\}}$.

	\item Each internal vertex $v$ (that is a vertex of the type $V$) corresponds to a graph $\Gamma\in\T G$, which is colored by a contractor $A_\Gamma$ (see \reef{A.Gamma}). This induces the coloring of the vertex: $A_v:=A_\Gamma$.
\end{itemize}
This completes the definition of the coloring.


\section{Examples of diagrams\label{sc:examples}}

\subsection{The very first example}
The very first example of  operator spin network diagram  has already been presented in \sref{IntrExample} and motivated our definitions.

\subsection{The trivial (static) spin foams\label{sc:trifialSF}} 
A trivial operator spin foam is, briefly speaking, defined by the histories of  constant in time spin networks.  It is natural to ask what operator spin network diagram gives as the result a trivial spin foam. The question is somewhat tricky, because the way our framework was introduced was motivated by decomposing a foam into neighbourhoods
of internal vertices. The trivial spin foams, on the other hand, have no internal vertices. Therefore an answer will not be completely trivial. This example teaches us which elements of the diagrams should be thought of as the trivial evolution
(nothing happening, no ``interaction'').

Given $(\Gamma, \rho)$, that is an oriented graph labelled by representations, consider the operator spin foam representing the trivial evolution. The corresponding foam has the topology $\kappa=\Gamma\times[0,1]$. The boundary graph is $\Gamma_\inn\cup\Gamma_\out$, where $\Gamma_\inn = \Gamma$ and $\Gamma_\out$ is obtained from  $\Gamma$ by switching the orientations of all the links.
For each link $\ell$ of $\Gamma_{\inn}$, the face $\ell\times [0,1]$ of $\kappa$ is oriented in the agreement with $\ell$ and colored by $\rho_\ell$. For each node $n$ of $\Gamma_{\inn}$, the corresponding internal edge $n\times [0,1]$ of the foam is
colored by the operator $P_n\in {\cal H}_n\otimes{\cal H}_n^*$ defined by the natural contraction (that is, $P_n$ defines the operator ${\rm id}: {\cal H}_n\rightarrow {\cal H}_n$). That data defines an operator spin foam $(\kappa,\rho,P)$ (due to the absence of  internal vertices, no vertex contractors are needed).   
Example of a foam $\kappa$ is shown at \fref{6_trivial_sf}.

\begin{figure}[hbt!]
	\centering
	\includegraphics[width=0.40\textwidth]{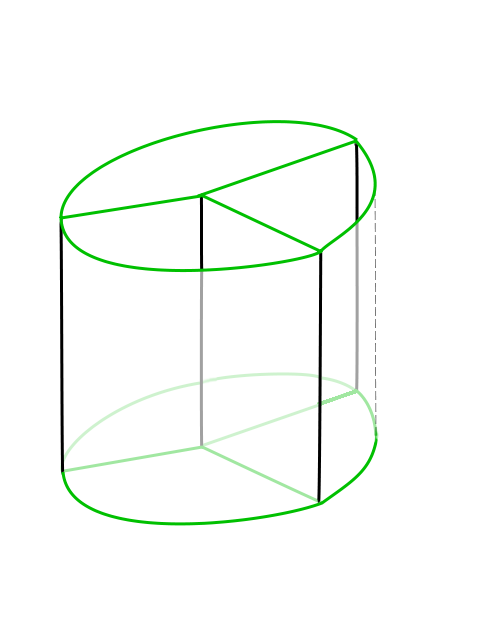}
	\caption{A trivial (static) spin foam}
	\label{fig:6_trivial_sf}
\end{figure}

We give now a receipt for an operator spin network diagram which gives an equivalent operator spin foam. The diagram will consist of so called generalised $\theta$-graphs. 

\begin{itemize}
	\item For each node $n$ of the  graph $\Gamma_\inn$ we introduce one graph $\tilde\theta_n$ in the following way (see also \fref{6_graph_to_theta}):
		\begin{enumerate}
			\item The graph $\theta_n$ is defined as follows. It has two nodes $n_\inn$ and $n_\out$. For each outgoing link $\ell$ at the node $n$ in $\Gamma$ there is one link $\ell^{(s)}$ at $n_\inn$ to $n_\out$ in $\theta_n$. For each incoming link $\ell$  at the node $n$ in $\Gamma$ there is one link $\ell^{(t)}$ going from $n_\out$ to $n_\inn$ in $\theta_n$.
			\item We construct the graph $\tilde\theta_n$ by adding a node at each link of $\theta_n$ (and splitting the link into two new links). Each new node will be denoted either by $s_\ell$ if it is on the link $\ell^{(s)}$ or by $t_\ell$ if it is on the link $\ell^{(t)}$. The new links will be denoted by $\ell^{(s/t)}_{\inn/\out}$ respectively (see \fref{6_graph_to_theta}). The new links inherit the orientation of the links of $\theta_n$.
		\end{enumerate}
\begin{figure}[hbt!]
	\centering
	\includegraphics[width=0.80\textwidth]{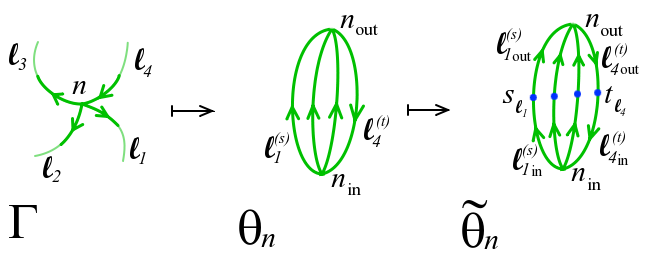}
	\caption{Construction of generalised $\theta$-graph from a node $n$.}
	\label{fig:6_graph_to_theta} 
\end{figure}
	
	\item For each link $\ell$ of the initial graph $\Gamma$ the node relation 
	$\T R_{\rm node}$ is defined to relate  
	the node $s_\ell$ of $\tilde\theta_{s(\ell)}$ and the node $t_\ell$ of $\tilde\theta_{t(\ell)}$. 
		
		At each pair $(s_\ell,t_\ell)$, the link relation $\T R^{(s_\ell,t_\ell)}_{\rm link}$ is defined to relate the link $\ell^{(s)}_\inn$ with $\ell^{(t)}_\inn$ and, respectively, the link $\ell^{(s)}_\out$ with $\ell^{(t)}_\out$ (i.e. it does not mix $\inn$- and $\out$-links).
		
		Note that no node of type $s_\ell$, or $t_\ell$ is left unrelated and all nodes $n_\inn$ and $n_\out$ \emph{are} unrelated (i.e. they are boundary nodes).
	
	\item We set the following coloring:
		\begin{enumerate}
			\item Each link of each $\tilde\theta$ graph is colored by the representation of the link of $\Gamma$ it comes from.
			\item Each boundary node $n$ and  each pair of the related internal nodes $\{n',n''\}$ is colored by the identity operator, the canonical element of the corresponding space ${\cal H}_n\otimes{\cal H}_n^*$, and respectively, of ${\cal H}_{n'}\otimes {\cal H}_{n''}$.   
			\item Each graph in the diagram is colored by the natural contractor $A^\tr$.
		\end{enumerate}
\end{itemize}
The \fref{6_triv_diag} shows the resulting graph diagram (the natural colorings 
are described above). Now, the foam defined by this graph diagram is not exactly
the trivial one \fref{6_trivial_sf}. Instead, we have obtained the spin foam presented at
\fref{6_trivial_sf_2}. It is obtained by dividing each of the faces of the original foam by a horizontal edge and extending the colorings in such a way, that
the resulting operator is unchanged. Hence, the foam we have obtained is equivalent to the trivial one.    

What we learn from this example is that the $\theta$ graphs colored by the canonical trace contractors and identity operators, accompanied with suitable node relations, play the role of identities (no interaction) in the spin network diagrams.

\begin{figure}[hbt!]
	\centering
	\subfloat[The graph diagram $\T D$ corresponding to the trivial spin foam. The dotted lines show the link relation. The node relation is omitted.]{\label{fig:6_trivial_diagram}\includegraphics[width=0.31\textwidth]{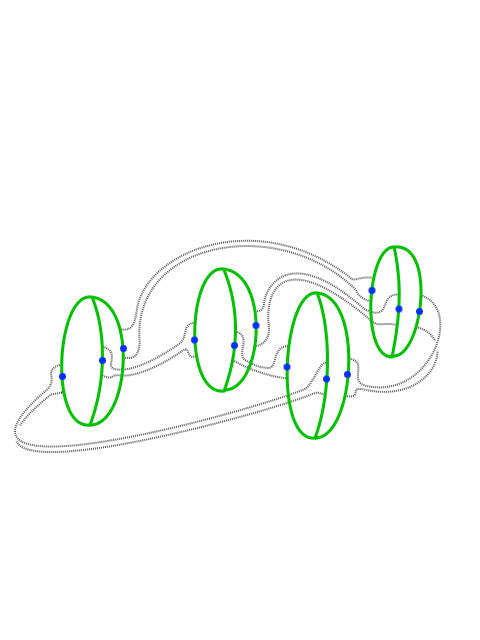}}
	\hspace{0.04\textwidth}
	\subfloat[The red (disjoint) graph is the boundary graph of $\T D$ (dashed red lines show the correspondence between nodes of the diagram and nodes of the boundary graph).]{\label{fig:6_trivial_w_boundary}\includegraphics[width=0.33\textwidth]{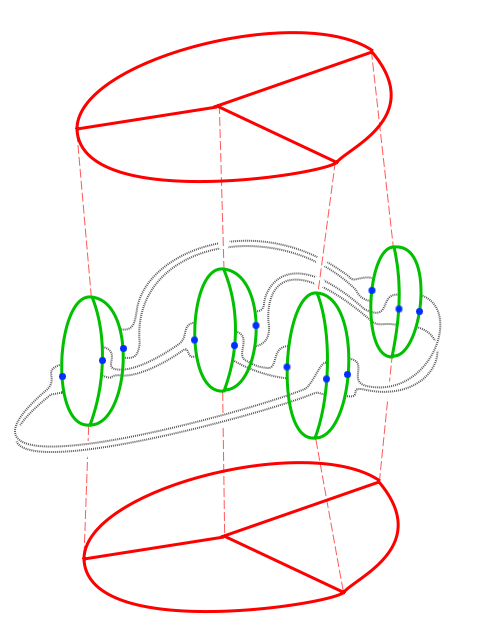}}
	\subfloat[The spin foam constructed from the diagram $\T D$. The \emph{horizontal} internal edges are all bivalent.]{\label{fig:6_trivial_sf_2}\includegraphics[width=0.31\textwidth]{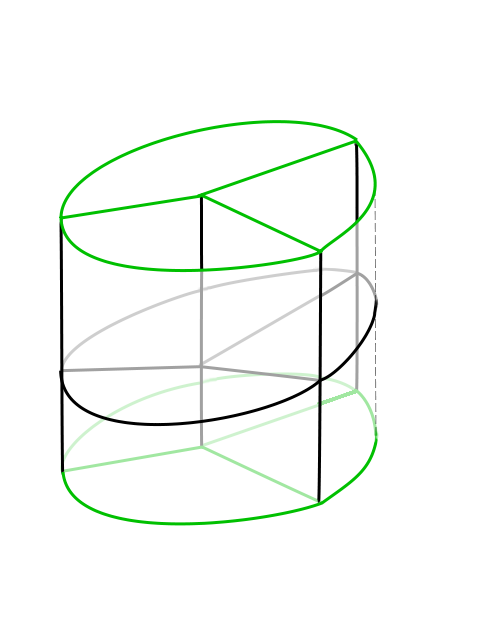}}
	\caption{The trivial graph diagram and reconstruction of corresponding 2-complex. The node relation is not drawn for the simplicity of the figure, but it can be read from the link relation at pairs of blue nodes.}
	\label{fig:6_triv_diag}
\end{figure}

\subsection{One interaction vertex spin foams\label{sc:interactions}}
Now we will use our formalism to describe a simple non-trivial evolution of a spin network. First we test the formalism on an very well known example of a foam.
Next, we show a quite simple diagram whose corresponding foam exceeds our graphical skills. 

Consider a one internal vertex  operator spin foam defining  the evolution of the spin network states
 on a graph $\Gamma_\inn$ whose links are colored by $\rho_\inn$ (with representations
of a group $G$) into the spin networks  on a graph $\Gamma_\out$ whose links are colored by $\rho_\out$.  Suppose for the simplicity, that all the operators coloring  the internal edges are the identities, and the internal vertex is colored by a contractor $A_v$. 

The neighborhood of the vertex is bounded by a graph $\Gamma_{\rm int}$ (see \fref{6_simple_to_graphs})
endowed with: the induced link coloring $\rho_{\rm int}$, node coloring $P_{\rm int}$, the contractor $A_{\rm int}=A_v$, and relating
some of its nodes with the initial graph, and the other nodes with the final graph.
This  information defines the  nontrivial evolution. The quadruple referred to as interaction operator spin network $(\Gamma_{\rm int},
\rho_{\int},P_{\rm int}, A_{\rm int})$ becomes an element of the corresponding operator spin network diagram 
(\fref{6_simple_diagram}).

\begin{figure}[hbt!]
	\centering
	\subfloat[The spin foam with one simple interaction vertex]{\label{fig:6_simple}\includegraphics[width=0.40\textwidth]{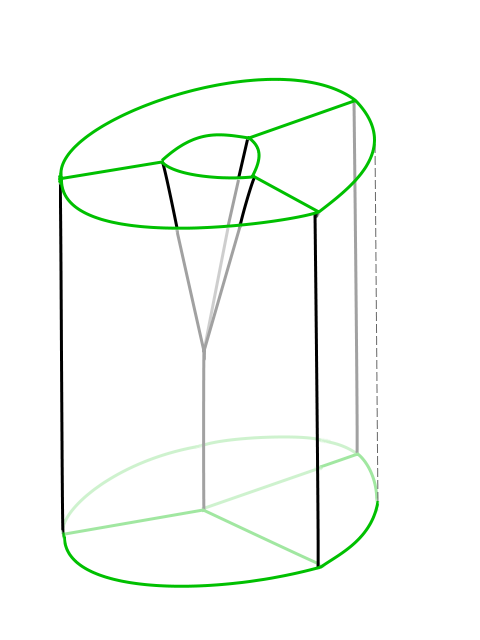}}
	\hspace{0.1\textwidth}
	\subfloat[The  interaction graph $\Gamma_{\rm int}$ (green), together with the boundary graphs $\Gamma_\inn$ and $\Gamma_\out$, respectively (red).]{\label{fig:6_simple_to_graphs}\includegraphics[width=0.40\textwidth]{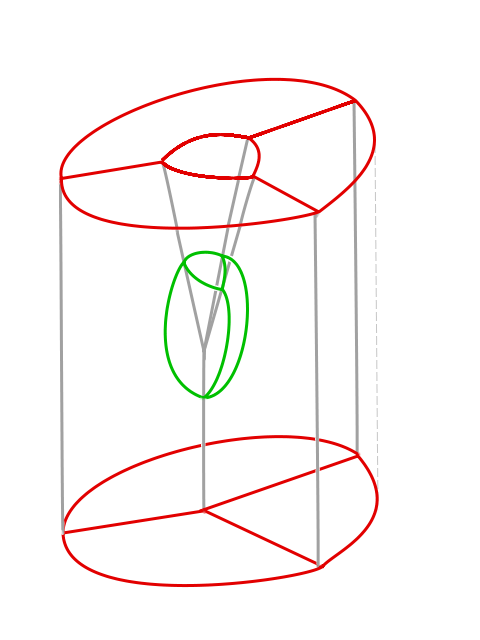}}
	\caption{A simple example of spin foam with one internal vertex.}
\end{figure}

To construct an operator spin network diagram representing this operator spin foam we first perform the construction of the previous example to the initial data
$(\Gamma_\inn, \rho_\inn)$. The result of this intermediate step is the operator spin network diagram  of the previous example.  Next, we extend it by the interaction operator spin network $(\Gamma_{\rm int},\rho_{\rm int},P_{\rm int}, A_{\rm int})$. The relation $\T R$ is extended in the way depicted at (see \fref{6_simple_diagram}). 

\begin{figure}[hbt!]
	\centering
	\subfloat[The graph diagram corresponding to the spin foam at \fref{6_simple} (the node relation is omitted)]{\label{fig:6_simple_diagram}\includegraphics[width=0.40\textwidth]{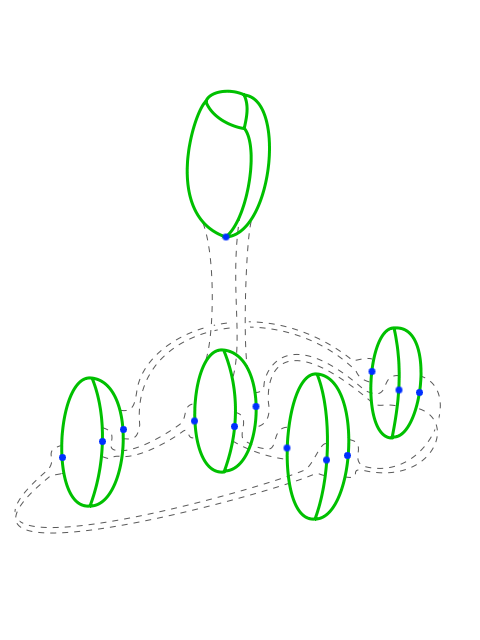}}
	\hspace{0.1\textwidth}
	\subfloat[The spin foam obtained from the graph diagram. The extra ''horizontal'' internal edges are trivial]{\label{fig:6_simple_of_diagram}\includegraphics[width=0.40\textwidth]{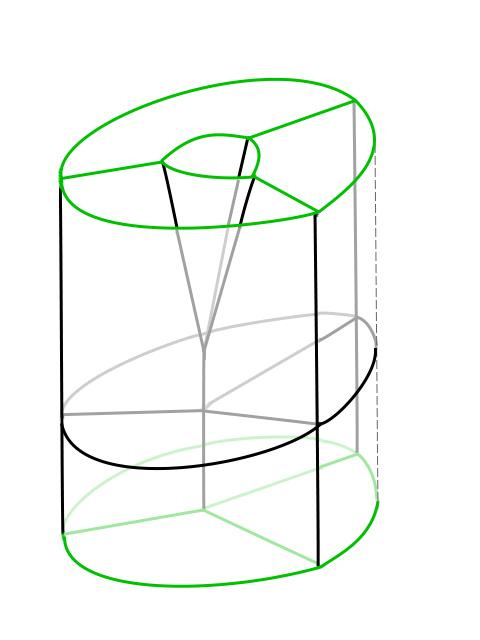}}
	\caption{The diagrammatic construction of the spin foam presented above.}
\end{figure}

The above example uses a very simple form of spin foam. We choose it because it is easy to draw the corresponding 2-complex explaining the construction. However, the power of diagrammatic formalism sits in more complicated diagrams, when drawing the spin foam on a 2-dimensional sheet of paper is difficult or even impossible. Consider a  graph diagram shown at \fref{6_compl}. For every coloring turning this graph into an operator spin network diagram, the calculation of the corresponding operator defined for  the boundary graph  (\fref{6_compl_bound}) is quite simple.

\begin{figure}[hbt!]
	\centering
	\subfloat[A graph diagram. The node relation between $\theta$ graphs is omitted. The link relation between links of $\theta$-graphs and interaction graph is described by the letters, i.e. $a$ is in relation with $a'$, $b$ with $b'$ etc.]{\label{fig:6_compl}\includegraphics[width=0.40\textwidth]{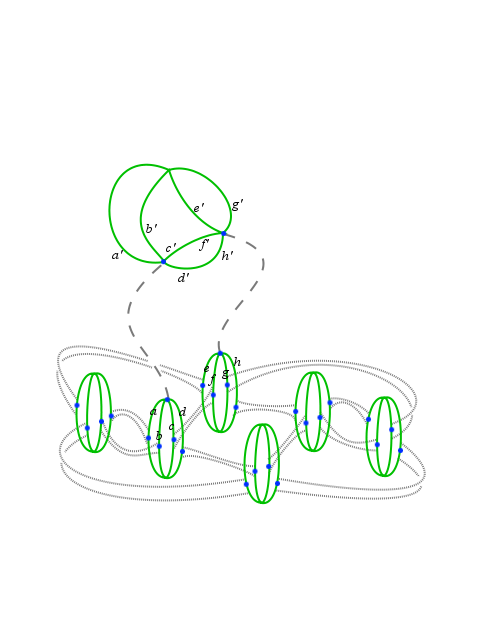}}
	\hspace{0.1\textwidth}
	\subfloat[The boundary in and out graphs (red) corresponding to the diagram (green). The red dashed lines show the correspondence between boundary nodes and diagram nodes.]{\label{fig:6_compl_bound}\includegraphics[width=0.40\textwidth]{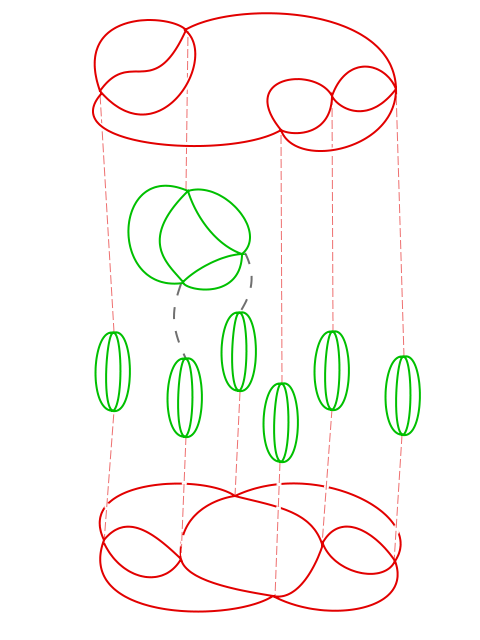}}
	\caption{A more complicated graph diagram.}
\end{figure}


\section{Summary, conclusions and outlook}
The  operator spin network diagrams and their framework is suited to play the analogous role in the covariant formulations of LQG to the Feynman diagrams in QFT. Our diagram description provides an itemisation of the  operator spin foams in terms of simpler elements: graphs,  node/link relations and colors. Similar ideas were introduced before by Frank Hellmann in his PhD thesis \cite{FrankPhD}.

The diagram framework introduced in this paper is capable enough to include the EPRL
spin foam model of the boundary Hilbert space equal the LQG kinematical Hilbert space and of either the Euclidean or the Lorentzian signature. Also, the natural operator spin network models introduced in \cite{Operator_SF} can be equivalently described
by another class of the operators spin network diagrams.   

There are two ways of thinking of the spin foam models of gravity.
   
The first one is orthodox covariant, in which the states of the theory are defined on
spin foam boundary.  It admits a natural formulation in terms of the operator spin network diagrams presented  in \sref{ExamplesOfApplication}.   
    
The second one splits the boundary into the initial and final parts
supporting the initial, and respectively, final states. 
The application of the diagram framework to the  initial/final state transition amplitudes was addressed in \sref{examples}. From those examples a scheme
of a theory defined by the operator spin network diagrams emerges.    
A specific theory can be defined by using the following elements: 
\begin{itemize}
\item a fixed set of the {\it interaction} graphs of the links colored by representations, nodes by operators and themselves colored by contractors 
\item the set of the ``propagators'' (the trivial interaction graphs), that is the generalized theta graphs constructed in \sref{trifialSF} of the links colored by the group representations, nodes colored by the identity operators and themselves colored by the natural trace contractors. 
\end{itemize}
With these blocks we first construct all the possible 1-interaction vertex diagrams,
and next all their compositions. 

Suppose, that the operators coloring the nodes of the diagrams are restricted to be projections only. Then, each colored graph in an operator spin network diagram can be assigned an operator on its own in such a way, that  the spin network diagram operator becomes the composition of the vertex and propagator operators. That farther
simplifies the framework.    

There are several technical problems we have not addressed in this paper but we will
do it elsewhere.  We briefly discuss them now. 

We claim that the 2-complexes obtained from the graph diagrams set the right class of the 2-complexes for the spin foams models of LQG to be defined on. The first question is whether there are foams that can not be obtained in this way. More exactly, what are the CW-complexes that are out of range by composing the graph diagrams? There are obvious degenerate examples in which a vertex or an edge is intersected by no face, but those are not used for foams. Are there any proper examples?  

The second question concerns equivalence between different diagrams. Certainly there are differently looking diagrams which define the same operators. For example the diagrams i \sref{interactions} are written in a way breaking the time symmetry. It is not hard 
to first restore the symmetry by adding on the top one more diagram 
representing the static foam of the final state. Next, the lower static foam diagram
(corresponding to the initial state) can be in a suitable way  removed.
The resulting diagram is equivalent but looks differently.  Another source
of the equivalent diagrams is  the spin network cylindrical consistency equivalence.

In the technical part of the construction of a 2-complex from a graph diagram the squid graphs were introduced. Their usefulness suggests they may play more important role than an auxiliary tool. Do they play a fundamental role by any chance?

One of the open problems of the spin foam approaches to the 4D gravity is
definition of the total amplitude that takes into account all the foams.
A recent breakthrough in this issue is  Rovelli-Smerlak's projective limit
definition \cite{Rovelli_Smerlak}. How do our diagrams fit in this limit?
 
Those questions will be answered soon either by us or by the readers.

\section*{Acknowledgments}
We thank Wojtek Kami\'nski for very useful remarks, in particular for the ideas of the: simplified notation for the interaction graph diagrams and contractors coloring the graphs. We also thank Benjamin Bahr, Frank Hellmann, William Nelson and Carlo Rovelli for their comments. Marcin Kisielowski and Jacek Puchta acknowledges financial support from the project "International PhD Studies in Fundamental Problems of  Quantum Gravity and Quantum Field Theory" of Foundation for Polish Science, cofinanced from the programme IE OP 2007-2013 within European Regional Development Fund. The work was also partially supported by the grants N N202 104838, and 182/N-QGG/2008/0 (PMN) of Polish Ministerstwo Nauki i Szkolnictwa Wy\.zszego. All the authors benefited from the travel grant of the ESF network Quantum Geometry and Quantum Gravity.

\appendix

\section{$\Delta$-complexes\label{sc:delta}}

To make the paper self contained we will provide here a definition of $\Delta$-complex. We will also give the strict definition in terms of $2-\Delta$-complexes of the edge-gluing procedure, which is the base of the gluing procedure used in the paper. Finally we will prove the theorem saying that (under some assumptions) gluing along two pairs of edges commute (we are not sure whether it is the strongest version of the theorem, however it is sufficient for our needs).

In our considerations \emph{$n$-simplex} will always mean \emph{$n$-simplex with ordered vertices}. A $n$-simplex will be denoted by $\Delta^n$. While considering 2-dimensional complexes we will use $\Delta$ without superscript to denote a two-simplex. One-simplexes will be called \emph{intervals} when considered separately (and denoted then by $I$), \emph{edges} when embedded into a $2-\Delta$-complex, and \emph{links} when considered as elements of boundary complex (i.e. graph). The zero-simplexes will be called \emph{vertices}, when considered as elements of $2$-complex, and \emph{nodes}, when they are elements of graph, and denoted $v$ and $n$ respectively.

\subsection{The definition}
Consider a number of sets $\T C_m$, where $m\in\{0,1,\ldots,n\}$, each of them containing $m$-simplexes: $\T C_m=\{\Delta^m_1,\ldots,\Delta^m_{N_m}\}$ (number $N_m$ is not necessarily finite). For each of them one can define a boundary set $\pd \T C_m =\bigsqcup_{i=1}^{N_m}\pd\Delta^m_{i}$. The boundary of a $m$-simplex is always union of $(m-1)$-simplexes.

Now consider functions $f_{m\to m-1}$ for $m=0,\ldots, n-1$ and set of relations $\sim_0,\ldots,\sim_{n-1}$ , such that
\begin{itemize}
	\item each relation $\sim_m$ is defined on the set $\T C_m$
	\item relation $\sim_0$ is the identity relation
	\item each function $f_{m\to m-1}$ is a map $f_{m\to m-1}:\pd\T C_m\to\T C_{m-1}/_{\sim_{m-1}}$
	\item the next relation $\sim_m$ is defined by the function $f_{m\to m-1}$ by:
		\be
			x\sim_m y \Leftrightarrow f_{m\to m-1}(x)=f_{m\to m-1}(y)
		\ee
\end{itemize}
The function $f_{m\to m-1}$ are called the boundary functions and they define the way that higher-dimension simplexes are glued onto lower dimension skeleton. It is worth to notice that since first relation, $\sim_0$, is a trivial relation, it can be omitted in the construction. Then any other relation is inductively constructed from functions $f_{m\to m-1}$. Thus what is essential in the construction of $\Delta$-complex are the boundary functions, not the relations (however they are very useful in geometrical interpretation).

Having these notions we may define a $\Delta$-complex:
\begin{df}
	$\Delta$-complex is a collection of sets $\T C_i$, where $i=0,1,\ldots,n$ together with the functions $f_{i\to i-1}$ for $i=1,\ldots,n$ defined as above.
	\be
		\kappa=\left(\T C_n,\ldots,\T C_0\;;\;f_{n\to n-1},\ldots,f_{1\to0}\right)
	\ee
\end{df}

When considering $2-\Delta$-complexes we will use notation
\be
	\kappa=\left(\T F,\T E,\T V\;;\;f_{2\to1},f_{1\to0}\right)
\ee


\subsection{How to glue a $2-\Delta$-complex along a pair of edges?\label{sc:cwgluing}}

We will define now the procedure of identifying two edges in a $2-\Delta$-complex. The definition is a special case of such procedure, which can be given for arbitrary dimension of both the complex and the simplexes to glue.

\begin{df}{Gluing along two edges}
	Given a $2-\Delta$-complex $\kappa=(\T F, \T E, \T V\;;\;f_{2\to1},f_{1\to0})$ and a pair of (different) edges $\alpha=(e_A,e_B)$ of $\T E$ one may define a $2-\Delta$-complex $\kappa/\alpha$ being the complex $\kappa$ with the edges $e_A$ and $e_B$ glued together. The resulting complex has the form:
	\be
		\kappa/\alpha=\left(\T F,\T E/\alpha_1, \T V/\alpha_0\;;\;\pi_{\alpha_1}\circ f_{2\to1},\pi_{\alpha_0}\circ f_{1\to0}\circ\left(\pi_{\alpha_1}\right)^{-1}\right)
	\ee
\end{df}

To make the definition complete, we have to specify the symbols that appears in above formula.

The set $\T E/\alpha_1$ is simply the set $\T E$ with edges $e_A$ identified with $e_B$. Formally it can be written as
\be
	\T E/\alpha_1\ni[e]=\left\{
		\begin{array}{ccl}
			e&\Leftrightarrow& e\not\in\{e_A,e_B\}\\
			{[e_A]}&\Leftrightarrow& e\in\{e_A,e_B\}
		\end{array}
	\right.
\ee
where $[e_A]$ when considered combinatorially is a single element labeled by such label, and when considered topologically (as an edge) acts just as its representant (i.e. $[e_A](t)=e_A(t)$). The projection map $\pi_{\alpha_1}:\T E\to\T E/\alpha_1$ is obvious.

The set of vertices $\T V/\alpha_0$ is the set $\T V$ with ends of edges $e_A$ and $e_B$ appropriately identified. This procedure is intuitively obvious, however need some care when being defined formally.

Lets name the beginning vertex of $e_A$ by $v_{A0}$, its ending vertex by $v_{A1}$, and respectively $v_{B0}$ and $v_{B1}$ for $e_B$ (i.e. $f_{1\to0}(s(e_A))=:v_{A0}$ etc.).  If each of $v_{A0},v_{A1},v_{B0},v_{B1}$ is different vertex, then the quotient space $\T V/\alpha_0$ is as easy to construct, as in case of $\T E/\alpha_1$. However it is possible, that some (or even all) of vertices $v_{A0},\ldots,v_{B1}$ are the same. We will consider two cases: first when in the resulting quotient space there is one equivalence class for all of that points, and second when there are two equivalence classes for them (only the later one were used in the paper).

The first case arises when at leas one of the following equalities holds:
\be\label{eq:cond.v}
	v_{A0}=v_{A1} \tab{\rm or}\tab v_{A1}=v_{B0}
\ee
or any of those two with $A$ and $B$ replaced\footnote{Since all the procedure is symmetric with respect to change of $e_A$ and $e_B$, any consequent change of $A$ and $B$ makes all the statements valid.}. In such case the two edges are mapped to a circle with one vertex on it, and the quotient vertex space is 
\be
	\T V/\alpha_1\ni[v]=\left\{
		\begin{array}{ccl}
			v&\Leftrightarrow& v\not\in\{v_{A0},v_{A1},v_{B0},v_{B1}\}\\
			{[v_{A0}]}&\Leftrightarrow& v\not\in\{v_{A0},v_{A1},v_{B0},v_{B1}\}
		\end{array}
	\right.
\ee
If none of conditions (\ref{eq:cond.v}) is satisfied (i.e edges either do not intersect or intersect at their beginnings or endings, or both, but ending with ending and beginning with beginning), the result of gluing is not a circle, but an interval, and the quotient vertex space is
\be
	\T V/\alpha_1\ni[v]=\left\{
		\begin{array}{ccl}
			v&\Leftrightarrow& v\not\in\{v_{A0},v_{A1},v_{B0},v_{B1}\}\\
			{[v_{A0}]}&\Leftrightarrow& v\not\in\{v_{A0},v_{B0}\}\\
			{[v_{A1}]}&\Leftrightarrow& v\not\in\{v_{A1},v_{B1}\}
		\end{array}
	\right.
\ee
The action of the projection map $\pi_{\alpha_0}$ in both cases is obvious.

What one should note is that in spite of presence of $\left(\pi_{\alpha_1}\right)^{-1}$ in the boundary function $\pi_{\alpha_0}\circ f_{1\to0}\circ\left(\pi_{\alpha_0}\right)^{-1}$, the boundary function is well defined, there is only one case, when $\left(\pi_{\alpha_1}\right)^{-1}$ is multivalued ($[e_A]$), and in that case $\pi_{\alpha_0}\circ f_{1\to0}$ gives the same result for both $e_A$ and $e_B$.

\subsection{Theorem of commutativity\label{sc:theorem}}
In our paper a certain special case of the gluing procedure is performed. All the edges we glue are boundary edges. And since the boundary of one-vertex-spinfoams are squid-graphs (see sec.\ref{sc:1vfoam}), they have some very useful feature: all the boundary vertices may be divided into two types: \emph{i)} these which have only outgoing boundary edges (heads of the squids), and \emph{ii)} those, which have only ingoing boundary edges (leg-nodes).

Since only the boundary edges are glued, this feature provides that only gluing of the second type appears, i.e. it is not possible to glue two edges such that ending of one of them is the beginning of another. The feature holds during gluing of boundary edges, because after each gluing the boundary of new complex is subgraph of the original boundary.

Thanks to that fact it is sufficient for our use to sate and prove the theorem of commutativity under following assumption: consider four different edges grouped in two pairs $\alpha=(e_A,e_B)$ and $\beta=(e_C,e_D)$ such, that
\be\label{eq:assumption}
	\forall_{i,j=A,B,C,D}\;s(e_i)\not=t(e_j)
\ee

The commutativity theorem says
\begin{thm}
	For any $2-\Delta$-complex $\kappa$ and any four different edges $e_A,\ldots,e_D$ such, that (\ref{eq:assumption}) holds, the following identity is true
	\be
		\left(\kappa/\alpha\right)/\beta=\left(\kappa/\beta\right)/\alpha
	\ee
	where $\alpha=(e_A,e_B)$ and $\beta=(e_C,e_D)$.
\end{thm}
\emph{Proof:}

One should prove, that each part of the $2-\Delta$-complexes are equal.

The regime of \emph{faces} is trivial, since the gluing does not effect the set $\T F$.

The regime of $\emph{edges}$ is not trivial, by it is obvious. Since $\T E/\alpha=\left(\T E\setminus\{e_A,e_B\} \right)\cup\left\{[e_A]\right\}$ and since $\left\{e_A,e_B\right\}\cap\left\{e_C,e_D\right\}=\emptyset$, we have
\be
	\left(\T E/\alpha\right)/\beta=\left(\left(\left(\T E\setminus\{e_A,e_B\}\right)
			\cup\left\{[e_A]\right\}\right)\setminus\left\{e_C,e_D\right\}\right)
			\cup\left\{[e_C]\right\}
		=\left(\T E\setminus\{e_A,e_B,e_C,e_D\}\right)\cup\{[e_A],[e_C]\}
\ee
which is symmetric with respect to change of order of $\alpha$ and $\beta$.

Having the set equality $\left(\T E/\alpha\right)/\beta=\left(\T E/\beta\right)/\alpha$ one may consider action of the projection maps $\pi_{\beta_1}\circ\pi_{\alpha_1}$ and $\pi_{\alpha_1}\circ\pi_{\beta_1}$, which is obviously the same.

Now we may go to the \emph{vertexes} regime.

Thanks to the assumption (\ref{eq:assumption}) we may decompose the set $\T V$ into a disjoint sum
\be
	\T V=\T V_0\sqcup\T V_1\sqcup\T V_{\rm rest}
\ee
where $\T V_0=\{v_{A0},v_{B0},v_{C0},v_{D0}\}$ are the starting points of the glued edges, $\T V_1$ are respectively their ending points and $\T V_{\rm rest}=\T V\setminus\left(\T V_0\cup\T V_1\right)$. None of gluing act on $\T V_{\rm rest}$, and the action of gluing procedure on $\T V_0$ and $\T V_1$ is independent and may be considered separately.

Lets take a look on $\T V_0$. The first quotient can be noted as
\ba
	\T V_0/\alpha&=&\left(\T V_0\setminus\{v_{A0},v_{B0}\}\right)\cup\{[v_{A0}]\}
		=\left(\{v_{A0},v_{B0},v_{C0},v_{D0}\}\setminus\{v_{A0},v_{B0}\}\right)\cup\{[v_{A0}]\}
		\nonumber\\
		&=&\left(\{v_{C0},v_{D0}\}\setminus\{v_{A0},v_{B0}\}\right)\cup\{[v_{A0}]\}
\ea
where one cannot omit the subtraction in the $\left(\{v_{C0},v_{D0}\}\setminus\{v_{A0},v_{B0}\}\right)$ term, because we do not know whether the two sets intersect or not.

Now lets take the second quotient. Note that one does not identify now points $v_{C0}$ and $v_{D0}$, but their equivalence classes $[v_{C0}]$ and $[v_{D0}]$ with respect to the relation $\sim_\alpha$. The quotient is
\ba
	\left(\T V_0/\alpha\right)/\beta
		&=& \left( \left(\left(\{v_{C0},v_{D0}\}\setminus\{v_{A0},v_{B0}\}\right)
				\cup\{[v_{A0}]\}\right)\setminus\{[v_{C0}],[v_{D0}]\}\right)\cup\{[[v_{C0}]]\}\\
		&=& \left(\left(\{v_{C0},v_{D0}\}\setminus\{v_{A0},v_{B0}\}\right)
						\setminus\{[v_{C0}],[v_{D0}]\}\right)
				\cup\left(\{[v_{A0}]\}\setminus\{[v_{C0}],[v_{D0}]\}\right)\cup\{[[v_{C0}]]\}
				\nonumber
\ea
Now: if $\{v_{C0},v_{D0}\}\cap\{v_{A0},v_{B0}\}=\emptyset$, then the equivalence classes $[v_{C0}]$,$[v_{D0}]$ are just the elements $v_{C0}$ and $v_{D0}$. So in this case the first term gives the empty set, while in the second term the subtraction gives just $\{[v_{A0}]\}$, so finally the result set is $\{[v_{A0}],[v_{C0}]\}$. However if at least one of the later points (say $v_{C0}$) belongs to $\{v_{A0},v_{B0}\}$, then $[v_{C0}]=[v_{A0}]$, so the second term vanishes, and the first term is $\left(\{v_{D0}\}\setminus\{v_{A0},v_{B0}\}\right)\setminus\{[v_{D0}]\}$, which also vanishes: either because $[v_{D0}]=v_{D0}$ (which occurs for $v_{D0}\not\in\{v_{A0},v_{B0}\}$) or because $v_{D0}\in\{v_{A0},v_{B0}\}$. So finally the result set in the second case is $\left(\T V_0/\alpha\right)/\beta=\left\{\left[[v_{C0}]\right]\right\}$, which is equal to $\left\{\left[[v_{A0}]\right]\right\}$ (because $[v_{C0}]=[v_{A0}]$).

In both cases the set $\left(\T V_0/\alpha\right)/\beta$:
\be\label{eq:sedcrition.V0}
	\left(\T V_0/\alpha\right)/\beta=\left\{
		\begin{array}{ccl}
			\left\{[v_{A0}],[v_{C0}]\right\}&\;{\rm for}\;&\{v_{A0},v_{B0}\}\cap\{v_{C0},v_{D0}\}=\emptyset\\
			\left\{\left[[v_{C0}]\right]\right\}&\;{\rm for}\;&\{v_{A0},v_{B0}\}\cap\{v_{C0},v_{D0}\}\not=\emptyset
		\end{array}
	\right.
\ee
is insensitive for change of the order $\alpha$ and $\beta$, which was the object of the proof.

The same reasoning goes for the set $\T V_1$, and thus for all the set $\T V$

Since the the set $\left(\T V/\alpha\right)/\beta$ being the image of $\pi_{\beta_0}\circ\pi_{\alpha_0}$ is the same as the image of $\pi_{\alpha_0}\circ\pi_{\beta_0}$, it is reasonable to ask whether they are the same maps. The answer is in affirmative what obviously follows from the formula (\ref{eq:sedcrition.V0}) describing the set $\T V$.

\emph{Quod erat demonstrandum}.

For our use the following further consideration is needed: since we glue the series of pairs of edges $\alpha_1,\ldots,\alpha_k$, we need to know whether any reordering of this series is equivalent. However since any permutation can be composed out of transpositions of neighbour elements, the theorem of this section implies that any permutation of $\alpha$s gives the same quotient complex.


\begin{thebibliography}{99}
\bibitem{RR}
	{Reisenberger~MP}~(1994)~\emph{World sheet formulations of gauge theories and gravity}, {} {(\emph{Preprit} arXiv:gr-qc/9412035)}\\
	Reisenberger~MP, Rovelli~C~(1997)~\emph{``Sum over Surfaces'' form of Loop Quantum Gravity}, {Phys.Rev.}{\bf D56},{3490-3508} {(\emph{Preprit} arXiv:gr-qc/9612035v)}
\bibitem{Markopoulou}
	{Markopoulou~F}~(1997)~\emph{Dual formulation of spin-network evolution}, {} {(\emph{Preprit} arXiv:gr-qc/9704013)}
\bibitem{Baezintro}
	{Baez~J}~(2000)~\emph{An introduction to Spinfoam Models of BF Theory and Quantum Gravity}, {Lect.Notes Phys.}{\bf 543},{25-94} {(\emph{Preprit} arXiv:gr-qc/9905087v1)}
\bibitem{perez}
	{Perez~A}~(2003)~\emph{Spinfoam models for Quantum Gravity}, {Class.Quant.Grav.}{\bf 20},{R43} {(\emph{Preprit} arXiv:gr-qc/0301113v2)}
\bibitem{ThiemannSpinFoams}
	{Baratin~A, Flori~C, Thiemann~T}~(2009)~\emph{The Holst Spin Foam Model via Cubulations}, {} {(\emph{Preprit} arXiv:gr-qc/0812.4055v2)}\\
	{Muxin~Han , Thiemann~T}~(2009)~\emph{On the Relation between Operator Constraint --, Master Constraint --, Reduced Phase Space --, and Path Integral Quantisation}, {} {(\emph{Preprit} arXiv:0911.3428v1)}\\
	{Engle~J, Muxin~Han, Thiemann~T}~(2009)~\emph{Canonical path integral measures for Holst and Plebanski gravity. I. Reduced Phase Space Derivation}, {} {(\emph{Preprit} arXiv:0911.3433v1)}
\bibitem{Rovellibook}
	{Rovelli~C}~(2004)~\emph{Quantum Gravity}, {(Cambridge: Cambridge University Press)}
\bibitem{NP}
	{Noui~K, Perez~A}~(2005)~\emph{Three dimensional loop quantum gravity: physical scalar product and spin-foam models}, {Class.Quant.Grav.}{\bf 22},{1739-1762}
\bibitem{EPRL}
	{Engle~J, Livine~E, Pereira~R, Rovelli~C}~(2008)~\emph{LQG vertex with finite Immirzi parameter}, {Nucl.Phys.}{\bf B799},{136-149} {(\emph{Preprit} arXiv:0711.0146v2)}
\bibitem{flipped}
	{Engle~J, Pereira~R, Rovelli~C}~(2008)~\emph{Flipped spinfoam vertex and loop gravity}, {Nucl.Phys.}{\bf B798},{251-290} {(\emph{Preprit} arXiv:0708.1236v1)}\\
	{Livine~ER, Speziale~S}~(2007)~\emph{A new spinfoam vertex for quantum gravity}, {Phys.Rev.}{\bf D76},{084028} {(\emph{Preprit} arXiv:0705.0674v2)}
\bibitem{FK}
	{Freidel~L, Krasnov~K}~(2008)~\emph{A New Spin Foam Model for 4d Gravity}, {Class.Quant.Grav.}{\bf 25},{125018} {(\emph{Preprit} arXiv:0708.1595v2)}\\
	{Livine~ER, Speziale~S}~(2008)~\emph{Consistently Solving the Simplicity Constraints for Spinfoam Quantum Gravity}, {Europhys.Lett.}{\bf 81},{50004} {(\emph{Preprit} arXiv:0708.1915)}\\
	{Bojowald~M, Perez~A}~(2010)~\emph{Spin foam quantization and anomalies}, {Gen.Rel.Grav.}{\bf 42},{877-907} {(\emph{Preprit} arXiv:gr-qc/0303026)}
\bibitem{ASYMP1}
	{Barett~J, Dowdall~R, Fairbairn~W, Gomez~H, Hellmann~F}~(2009)~\emph{Asymptotic analysis of the EPRL four-simplex amplitude}, {J.Math.Phys.}{\bf 50},{112504} {(\emph{Preprit} arXiv:0902.1170)}
\bibitem{depietri} 
	{De Pietri~R, Petronio~C}~(2000)~\emph{Feynman Diagrams of Generalized Matrix Models and the Associated Manifolds in Dimension 4}, {J.Math.Phys.}{\bf 41},{6671-6688} {(\emph{Preprit} arXiv:gr-qc/0004045)} 
\bibitem{FreidelGFT}
	{Feidel~L}~(2005)~\emph{Group Field Theory: An overview}, {Int.J.Theor.Phys.}{\bf 44},{1769-1783} {(\emph{Preprit} arXiv:hep-th/0505016)}
\bibitem{Geloun:2010vj}
	{Geloun~JB, Gurau~R, Rivasseau~V}~(2010)~\emph{EPRL/FK Group Field Theory}, {}{(\emph{Preprit} arXiv:1008.0354)}
\bibitem{SFLQG}
	{Kami\'nski~W, Kisielowski~M, Lewandowski~J}~(2010)~\emph{Spin-Foams for All Loop Quantum Gravity}, {Class.Quantum~Grav.}{\bf 27},{095006} {(\emph{Preprit} arXiv:0909.0939v2)}
\bibitem{Benjamin}
	{Bahr~B}~(2001)~\emph{On knottings in the physical Hilbert space of LQG as given by the EPRL model}, {}{(\emph{Preprit} arXiv:1006.0700)}
\bibitem{LQGdiscr}
	{Rovelli~C, Smolin~C}~(1995)~\emph{Discreteness of area and volume in quantum gravity}, {Nucl.Phys.}{\bf B442},{593 [(1995) Erratum-ibid. \textbf{B456}  753]}  {(\emph{Preprit} arXiv:gr-qc/9411005)}\\
	{Ashtekar~A, Lewandowski~J}~(1995)~\emph{Differential Geometry on the Space of Connections via Graphs and Projective Limits}, {J.Geom.Phys.}{\bf 17},{191-230} {(\emph{Preprit} arXiv:hep-th/9412073)}\\
	{Ashtekar~A, Lewandowski~J}~(1997)~\emph{Quantum theory of geometry. I: Area operators}, {Class.Quant.Grav.}{\bf 14},{A55} {(\emph{Preprit} arXiv:gr-qc/9602046)}\\
	{Ashtekar~A, Lewandowski~J}~(1998)~\emph{Quantum theory of geometry. II: Volume operators}, {Adv.Theor.Math.Phys.}{\bf 1},{388} {(\emph{Preprit} arXiv:gr-qc/9711031)}\\ 
	{Thiemann~T}~(1998)~\emph{A length operator for canonical quantum gravity}, {J.Math.Phys.}{\bf 39},{3372} {(\emph{Preprit} arXiv:gr-qc/9606092)}
\bibitem{AshLewrev}
	{Ashtekar~A, Lewandowski~J}~(2004)~\emph{Background independent quantum gravity: A status report}, {Class.Quant.Grav.}{\bf 21},{R53} {(\emph{Preprit} arXiv:gr-qc/0404018)}
\bibitem{Marev}
	{Muxin~Han, Weiming~Huang, Yongge~Ma}~(2007)~\emph{Fundamental Structure of Loop Quantum Gravity}, {Int.J.Mod.Phys.}{\bf D16},{1397-1474} {(\emph{Preprit} arXiv:gr-qc/0509064)}
\bibitem{Ashtekarbook}
	{Ashtekar~A}~(1991)~\emph{Lectures on Non-perturbative Canonical Gravity}, {(Notes prepared in collaboration with R.S. Tate), (World Scientific Singapore)}
\bibitem{Thiemannbook}
	{Thiemann~T}~(2007)~\emph{Introduction to Modern Canonical Quantum General Relativity}, {(Cambridge: Cambridge University Press)}
\bibitem{SF_complex}
	{Oeckl~R}~(2001)~\emph{Generalized Lattice Gauge Theory, Spin Foams and State Sum Invariants}, {J.Geom.Phys.}{\bf 46},{308-35} {(\emph{Preprit} arXiv:hep-th/0110259)}\\
	{Oeckl~R}~(2005)~\emph{Discrete gauge theory: From Lattices to TQFT}, {Imperial College Press}
\bibitem{thiemann}
	{Baratin~A, Flori~C, Thiemann~T}~(2008)~\emph{The Holst Spin Foam Model via Cubulations}, {}{(\emph{Preprit} arXiv:arXiv:0812.4055v2)}
\bibitem{engle}
	{Engle~J}~(2008)~\emph{Piecewise linear loop quantum gravity}, {}{(\emph{Preprit} arXiv:gr-qc/0812.1270v1)}
\bibitem{FrankPhD}
	{Hellmann~F}~(2011)~\emph{State Sums and Geometry}, {PhD Thesis} {(\emph{Preprit} arXiv:1102.1688v1)}
\bibitem{cEPRL}
	{Kami\'nski~W, Kisielowski~M, Lewandowski~J}~(2010)~\emph{The EPRL intertwiners and corrected partition function}, {Class.Quant.Grav.}{\bf 27},{165020} {(\emph{Preprit} arXiv:0912.0540v1)}
\bibitem{Operator_SF}
	{Bahr~B, Hellmann~F, Kami\'nski~W, Kisielowski~M, Lewandowski~J}~(2010)~\emph{Operator Spin Foam Models}, {Class.Quant.Grav.}{\bf 28},{105003,2011} {(\emph{Preprit} arXiv:1010.4787v1)}
\bibitem{Rovelli_Smerlak}
	{Rovelli~C, Smerlak~M}~(2010)~\emph{In quantum gravity, summing is refining}, {}{(\emph{Preprit} arXiv:1010.5437v3)}
\bibitem{EPRL_contractor}
	{Rovelli~C}~(2010)~\emph{A new look at loop quantum gravity}, {Class.Quant.Grav.}{\bf 28},{114005,2011} {(\emph{Preprit} arXiv:1004.1780v4)}
\bibitem{EPRL_full}
	{Bianchi~E, Regoli~D, Rovelli~C}~(2010)~\emph{Face amplitude of spinfoam quantum gravity}, {Class.Quant.Grav.}{\bf 27},{185009,2010} {(\emph{Preprit} arXiv:1005.0764v1)}\\
	{You~Ding, Muxin~Han, Rovelli~C}~(2011)~\emph{Generalized Spinfoams}, {Phys.Rev.}{\bf D83},{124020,2011} {(\emph{Preprit} arXiv:1011.2149v2)}\\
	{Rovelli~C}~(2011)~\emph{Zakopane lectures on loop gravity}, {}{(\emph{Preprit} arXiv:1102.3660v3)}
\bibitem{Bianchi:2010mw}
	{Bianchi~E,Magliaro~E, Perini~C}~(2010)~\emph{Spinfoams in the holomorphic representation}, {Phys.Rev.}{\bf D82},{124031} {(\emph{Preprit} arXiv:1004.4550)}
\bibitem{SF_cosmology}
	{Bianchi~E, Rovelli~C, Vidotto~F}~(2010)~\emph{Towards Spinfoam Cosmology}, {Phys.Rev.}{\bf D82},{084035,2010} {(\emph{Preprit} arXiv:1003.3483v1)}
\bibitem{Kamykfinite}
	{Kami\'nski~W}~(2010)~\emph{All 3-edge-connected relativistic BC and EPRL spin-networks are integrable}, {}{(\emph{Preprit} arXiv:1010.5384v1)}
\bibitem{BC}
	{Barrett~JW, Crane~L}~(1998)~\emph{Relativistic spin-networks and quantum gravity}, {J.Math.Phys.}{\bf 39},{3296-3302} {(\emph{Preprit} arXiv:gr-qc/9709028)}
\bibitem{graviton}
	{Bianchi~E, Modesto~L, Rovelli~C, Speziale~S}~(2006)~\emph{Graviton propagator in loop quantum gravity}, {Class.Quant.Grav.}{\bf 23},{6989-7028} {(\emph{Preprit} arXiv:gr-qc/0604044)}\\
	{Alesci~E, Rovelli~C}~(2007)~\emph{The complete LQG propagator I. Difficulties with the Barrett-Crane vertex}, {Phys.Rev.}{\bf D76},{104012} {(\emph{Preprit} arXiv:gr-qc/0708.0883)}\\
	{Alesci~E, Rovelli~C}~(2007)~\emph{The complete LQG propagator: II. Asymptotic behavior of the vertex}, {Phys.Rev.}{\bf D77},{044024} {(\emph{Preprit} arXiv:0711.1284)}\\
	{Alesci~E, Bianchi~E, Rovelli~C}~(2008)~\emph{LQG propagator: III. The new vertex}, {}{(\emph{Preprit} arXiv:0812.5018)}\\
	{Mamone~D, Rovelli~C}~(2009)~\emph{Second-order amplitudes in loop quantum gravity}, {}{(\emph{Preprit} arXiv:0904.3730)}\\
	{Bianchi~E, Magliaro~E, Perini~C}~(2009)~\emph{LQG propagator from the new spin foams}, {Nucl.Phys.}{\bf B822},{245-269} {(\emph{Preprit} arXiv:0905.4082)}
\bibitem{BC_nvalent}
	{Yetter~D}~(1998)~\emph{Generalized Barrett-Crane vertices and invariants of embedded graphs}, {}{(\emph{Preprit} arXiv:math/9801131)}\\
	{Barrett~JW}~(1998)~\emph{The classical evaluation of relativistic spin networks}, {Adv.Theor.Math.Phys.}{\bf 2},{593-60} {(\emph{Preprit} arXiv:math/9803063)}
\bibitem{BC_R}
	{Reisenberger~MP}~(1999)~\emph{On relativistic spin network vertices}, {J.Math.Phys.}{\bf 40},{2046-2054} {(\emph{Preprit} arXiv:gr-qc/9809067v1)}
\bibitem{Han:2010rb}
	{Muxin~Han, Thiemann~T}~(2010)~\emph{Commuting Simplicity and Closure Constraints for 4D Spin Foam Models}, {}{(\emph{Preprit} arXiv:1010.5444)}
\bibitem{Bonzom:2008ru}
	{Bonzom~V, Livine~ER}~(2009)~\emph{A Lagrangian approach to the Barrett-Crane spin foam model}, {Phys.Rev.}{\bf D79},{064034} {(\emph{Preprit} arXiv:0812.3456)}\\
	{Bonzom~V}~(2009)~\emph{From lattice BF gauge theory to area-angle Regge calculus}, {Class.Quant.Grav.}{\bf 26},{155020} {(\emph{Preprit} arXiv:0903.0267)}\\
	{Bonzom~V}~(2009)~\emph{Spin foam models for quantum gravity from lattice path integrals}, {Phys.Rev.}{\bf D80},{064028} {(\emph{Preprit} arXiv:0905.1501)}
\bibitem{Dittrich:2008ar}
	{Dittrich~B, Ryan~JP}~(2008)~\emph{Phase space descriptions for simplicial 4d geometries}, {}{(\emph{Preprit} arXiv:0807.2806)}
	{Dittrich~B, Ryan~JP}~(2010)~\emph{Simplicity in simplicial phase space}, {Phys.Rev.}{\bf D82},{064026} {(\emph{Preprit} arXiv:1006.4295)}
\bibitem{Bahr:2009mc}
	{Bahr~B, Dittrich~B}~(2009)~\emph{Breaking and restoring of diffeomorphism symmetry in discrete gravity}, {}{(\emph{Preprit} arXiv:0909.5688)}
\bibitem{Krajewski:2010yq}
	{Krajewski~T, Magnen~J, Rivasseau~V, Tanasa~A, Vitale~P}~(2010)~\emph{Quantum Corrections in the Group Field Theory Formulation of the EPRL/FK Models}, {}{(\emph{Preprit} arXiv:1007.3150)}
\bibitem{BD}
	{Bahr~B, Dittrich~B}~(2009)~\emph{Regge calculus from a new angle}, {}{(\emph{Preprint} arXiv:gr-qc/0907.4325)}\\
	{Bahr~B, Dittrich~B}~(2009)~\emph{Improved and Perfect Actions in Discrete Gravity}, {}{(\emph{Preprint} arXiv:0907.4323)}
\bibitem{FC}
	{Conrady~F, Freidel~L}~(2009)~\emph{Quantum geometry from phase space reduction}, {}{(\emph{Preprint} arXiv:0902.0351)}
\bibitem{zapata}
	{Zapata~JA}~(2002)~\emph{Continuum spin-foam model for 3d gravity}, {J.Math.Phys.}{\bf 43},{5612-5623} {(\emph{Preprit} arXiv:gr-qc/0205037)}
\bibitem{pl} 
	{Rourke~C, Sanderson~B}~(1972)~\emph{Introduction to Piecewise-Linear Topology}, {(Springer Verlag, Berlin)} \\
	Lurie~J \emph{Topics in Geometric Topology}, [http://math.mit.edu/~lurie/937.html]
\bibitem{Hatcher}
	{Hatcher~A}~(2002)~\emph{Algebraic Topology}, {(Cambridge University Press)}
\bibitem{Holst}
	{Holst~S}~(1996)~\emph{Barbero's hamiltonian derived from a generalised Hilbert-Palatini action}, {Phys.Rev.}{\bf D53},{5966-5969}
\bibitem{barbero-immirzi}
	{Barbero~F}~(1996)~\emph{Real Ashtekar variables for Lorentzian signature space-times}, {Phys.Rev.}{\bf D51},{5507–5510}\\
	{Immirzi~G}~(1997)~\emph{Quantum gravity and Regge calculus}, {Nucl.Phys.Proc.Suppl.}{\bf 57},{65–72}
\bibitem{Reisenberger}
	{Reisenberger~MP}~(1997)~\emph{A left-handed simplicial action for euclidean general relativity}, {Class.Quant.Gravity}{\bf 14},{1753}{(\emph{Preprit} arXiv:gr-qc/9609002)}
\end{thebibliography}
\end{document}